\pdfoutput=1
\def\hlinew#1{\noalign{\ifnum0=`}\fi\hrule \@height #1 \futurelet\reserved@a\@xhline}
\documentclass[12pt,a4paper]{article}

\usepackage{ifthen} 
\newboolean{pdflatex}
\setboolean{pdflatex}{true} 

\newboolean{articletitles}
\setboolean{articletitles}{true} 

\newboolean{uprightparticles}
\setboolean{uprightparticles}{false} 

\def\paperauthors{LHCb collaboration} 
\def\paperasciititle{Search for the rare decay } 
\def\papertitle{Search for the rare decay $\Bd\to\jpsi\phi$} 
\def\paperkeywords{{High Energy Physics}, {LHCb}} 
\def\papercopyright{\the\year\ CERN for the benefit of the LHCb collaboration} 
\def\paperlicence{CC BY 4.0 licence}
\def\paperlicenceurl{https://creativecommons.org/licenses/by/4.0/}

\usepackage[top=1in, bottom=1.25in, left=1in, right=1in]{geometry}

\usepackage{microtype}
\usepackage{lineno}  
\usepackage{xspace} 
\usepackage{caption} 

\usepackage{graphicx}  
\usepackage{color}
\usepackage{colortbl}
\graphicspath{{./figs/}} 

\usepackage{amsmath} 
\usepackage{amssymb}
\usepackage{amsfonts}
\usepackage{upgreek} 

\newcommand*\patchAmsMathEnvironmentForLineno[1]{%
\expandafter\let\csname old#1\expandafter\endcsname\csname #1\endcsname
\expandafter\let\csname oldend#1\expandafter\endcsname\csname
end#1\endcsname
 \renewenvironment{#1}%
   {\linenomath\csname old#1\endcsname}%
   {\csname oldend#1\endcsname\endlinenomath}%
}
\newcommand*\patchBothAmsMathEnvironmentsForLineno[1]{%
  \patchAmsMathEnvironmentForLineno{#1}%
  \patchAmsMathEnvironmentForLineno{#1*}%
}
\AtBeginDocument{%
\patchBothAmsMathEnvironmentsForLineno{equation}%
\patchBothAmsMathEnvironmentsForLineno{align}%
\patchBothAmsMathEnvironmentsForLineno{flalign}%
\patchBothAmsMathEnvironmentsForLineno{alignat}%
\patchBothAmsMathEnvironmentsForLineno{gather}%
\patchBothAmsMathEnvironmentsForLineno{multline}%
\patchBothAmsMathEnvironmentsForLineno{eqnarray}%
}

\usepackage{hyperxmp}

\usepackage[pdftex,
            pdfauthor={\paperauthors},
            pdftitle={\paperasciititle},
            pdfkeywords={\paperkeywords},
            pdfcopyright={Copyright (C) \papercopyright},
            pdflicenseurl={\paperlicenceurl}]{hyperref}

\usepackage[colorinlistoftodos,textsize=scriptsize]{todonotes}

\usepackage[bottom,flushmargin,hang,multiple]{footmisc}

\usepackage[all]{hypcap} 

\usepackage{xspace} 
\usepackage{upgreek}

\def\lhcb   {\mbox{LHCb}\xspace}

\def\babar  {\mbox{BaBar}\xspace}
\def\belle  {\mbox{Belle}\xspace}

\def\MagUp {\mbox{\em Mag\kern -0.05em Up}\xspace}

\ifthenelse{\boolean{uprightparticles}}%
{

 \def\Pmu         {\ensuremath{\upmu}\xspace}                 
 \def\Pnu         {\ensuremath{\upnu}\xspace}                 
                  
 \def\Ppi         {\ensuremath{\uppi}\xspace}

 \def\Ppsi        {\ensuremath{\uppsi}\xspace}

 \def\PDelta      {\ensuremath{\Delta}\xspace}                 
 \def\PXi         {\ensuremath{\Xi}\xspace}                 
 \def\PLambda     {\ensuremath{\Lambda}\xspace}                 
 \def\PSigma      {\ensuremath{\Sigma}\xspace}                 
 \def\POmega      {\ensuremath{\Omega}\xspace}                 
 \def\PUpsilon    {\ensuremath{\Upsilon}\xspace}

 \def\PB      {\ensuremath{\mathrm{B}}\xspace}                 
                  
 \def\PD      {\ensuremath{\mathrm{D}}\xspace}

 \def\PJ      {\ensuremath{\mathrm{J}}\xspace}                 
 \def\PK      {\ensuremath{\mathrm{K}}\xspace}

 \def\Pb      {\ensuremath{\mathrm{b}}\xspace}                 
 \def\Pc      {\ensuremath{\mathrm{c}}\xspace}

 \def\Pi      {\ensuremath{\mathrm{i}}\xspace}

 \def\Pp      {\ensuremath{\mathrm{p}}\xspace}

 \def\Ps      {\ensuremath{\mathrm{s}}\xspace}

 \def\thebaroffset{0.0em}
}
{

 \def\Pmu         {\ensuremath{\mu}\xspace}                 
 \def\Pnu         {\ensuremath{\nu}\xspace}                 
                  
 \def\Ppi         {\ensuremath{\pi}\xspace}

 \def\Ppsi        {\ensuremath{\psi}\xspace}                 
                  
 \mathchardef\PDelta="7101
 \mathchardef\PXi="7104
 \mathchardef\PLambda="7103
 \mathchardef\PSigma="7106
 \mathchardef\POmega="710A
 \mathchardef\PUpsilon="7107
                  
 \def\PB      {\ensuremath{B}\xspace}                 
                  
 \def\PD      {\ensuremath{D}\xspace}

 \def\PJ      {\ensuremath{J}\xspace}                 
 \def\PK      {\ensuremath{K}\xspace}

 \def\Pb      {\ensuremath{b}\xspace}                 
 \def\Pc      {\ensuremath{c}\xspace}

 \def\Pi      {\ensuremath{i}\xspace}

 \def\Pp      {\ensuremath{p}\xspace}

 \def\Ps      {\ensuremath{s}\xspace}

 \def\thebaroffset{0.18em}
}
\newcommand{\offsetoverline}[2][\thebaroffset]{\kern #1\overline{\kern -#1 #2}}%

\makeatletter
\ifcase \@ptsize \relax
  \newcommand{\miniscule}{\@setfontsize\miniscule{4}{5}}
\or
  \newcommand{\miniscule}{\@setfontsize\miniscule{5}{6}}
\or
  \newcommand{\miniscule}{\@setfontsize\miniscule{5}{6}}
\fi
\makeatother

\DeclareRobustCommand{\optbar}[1]{\shortstack{{\miniscule (\rule[.5ex]{1.25em}{.18mm})}
  \\ [-.7ex] $#1$}}

\def\mup        {{\ensuremath{\Pmu^+}}\xspace}

\def\mumu       {{\ensuremath{\Pmu^+\Pmu^-}}\xspace}

\def\squark    {{\ensuremath{\Ps}}\xspace}

\def\cquark    {{\ensuremath{\Pc}}\xspace}

\def\bquark    {{\ensuremath{\Pb}}\xspace}

\def\pion   {{\ensuremath{\Ppi}}\xspace}
\def\piz    {{\ensuremath{\pion^0}}\xspace}
\def\pip    {{\ensuremath{\pion^+}}\xspace}
\def\pim    {{\ensuremath{\pion^-}}\xspace}
\def\pipm   {{\ensuremath{\pion^\pm}}\xspace}

\def\kaon    {{\ensuremath{\PK}}\xspace}
\def\Kbar    {{\ensuremath{\offsetoverline{\PK}}}\xspace}

\def\KorKbar {\kern \thebaroffset\optbar{\kern -\thebaroffset \PK}{}\xspace}
\def\Kz      {{\ensuremath{\kaon^0}}\xspace}

\def\Kp      {{\ensuremath{\kaon^+}}\xspace}
\def\Km      {{\ensuremath{\kaon^-}}\xspace}
\def\Kpm     {{\ensuremath{\kaon^\pm}}\xspace}

\def\Kstarz  {{\ensuremath{\kaon^{*0}}}\xspace}
\def\Kstarzb {{\ensuremath{\Kbar{}^{*0}}}\xspace}

\def\D       {{\ensuremath{\PD}}\xspace}

\def\DorDbar {\kern \thebaroffset\optbar{\kern -\thebaroffset \PD}\xspace}

\def\Dp      {{\ensuremath{\D^+}}\xspace}
\def\Dm      {{\ensuremath{\D^-}}\xspace}

\def\DpDm    {\ensuremath{\Dp {\kern -0.16em \Dm}}\xspace}

\def\B       {{\ensuremath{\PB}}\xspace}

\def\BorBbar {\kern \thebaroffset\optbar{\kern -\thebaroffset \PB}\xspace}

\def\Bd      {{\ensuremath{\B^0}}\xspace}

\def\BdorBdbar {\kern \thebaroffset\optbar{\kern -\thebaroffset \Bd}\xspace}
\def\Bu      {{\ensuremath{\B^+}}\xspace}

\def\Bs      {{\ensuremath{\B^0_\squark}}\xspace}

\def\BsorBsbar {\kern \thebaroffset\optbar{\kern -\thebaroffset \Bs}\xspace}
\def\Bc      {{\ensuremath{\B_\cquark^+}}\xspace}

\def\jpsi     {{\ensuremath{{\PJ\mskip -3mu/\mskip -2mu\Ppsi}}}\xspace}

\def\Y#1S{\ensuremath{\PUpsilon{(#1S)}}\xspace}

\def\proton      {{\ensuremath{\Pp}}\xspace}

\def\Lz          {{\ensuremath{\PLambda}}\xspace}

\def\LorLbar     {\kern \thebaroffset\optbar{\kern -\thebaroffset \PLambda}\xspace}

\def\Lb           {{\ensuremath{\Lz^0_\bquark}}\xspace}

\newcommand{\decay}[2]{\ensuremath{#1\!\to #2}\xspace} 

\def\to                 {\ensuremath{\rightarrow}\xspace}

\newcommand{\DGs}{{\ensuremath{\Delta\Gamma_{\squark}}}\xspace}

\newcommand{\Gs}{{\ensuremath{\Gamma_{\squark}}}\xspace}

\newcommand{\phis}{{\ensuremath{\phi_{\squark}}}\xspace}

\def\BsToJPsiPhi  {\decay{\Bs}{\jpsi\phi}}

\def\AT#1     {\ensuremath{A_{\mathrm{T}}^{#1}}\xspace}           

\def\C#1      {\ensuremath{\mathcal{C}_{#1}}\xspace}                       
\def\Cp#1     {\ensuremath{\mathcal{C}_{#1}^{'}}\xspace}                    
\def\Ceff#1   {\ensuremath{\mathcal{C}_{#1}^{\mathrm{(eff)}}}\xspace}        
\def\Cpeff#1  {\ensuremath{\mathcal{C}_{#1}^{'\mathrm{(eff)}}}\xspace}       
\def\Ope#1    {\ensuremath{\mathcal{O}_{#1}}\xspace}                       
\def\Opep#1   {\ensuremath{\mathcal{O}_{#1}^{'}}\xspace}                    

\newcommand{\nospaceunit}[1]{\ensuremath{\text{#1}}}       
\newcommand{\aunit}[1]{\ensuremath{\text{\,#1}}}       

\newcommand{\tev}{\aunit{Te\kern -0.1em V}\xspace}
\newcommand{\gev}{\aunit{Ge\kern -0.1em V}\xspace}
\newcommand{\mev}{\aunit{Me\kern -0.1em V}\xspace}
\newcommand{\kev}{\aunit{ke\kern -0.1em V}\xspace}
\newcommand{\ev}{\aunit{e\kern -0.1em V}\xspace}
 
\newcommand{\mevc}{\ensuremath{\aunit{Me\kern -0.1em V\!/}c}\xspace}
\newcommand{\gevc}{\ensuremath{\aunit{Ge\kern -0.1em V\!/}c}\xspace}
\newcommand{\mevcc}{\ensuremath{\aunit{Me\kern -0.1em V\!/}c^2}\xspace}
\newcommand{\gevcc}{\ensuremath{\aunit{Ge\kern -0.1em V\!/}c^2}\xspace}

\def\mum  {\ensuremath{\,\upmu\nospaceunit{m}}\xspace}

\def\fb   {\ensuremath{\aunit{fb}}\xspace}
\def\invfb   {\ensuremath{\fb^{-1}}\xspace}

\def\ps   {\ensuremath{\aunit{ps}}\xspace}

\newcommand{\chisq}{\ensuremath{\chi^2}\xspace}

\newcommand{\chisqip}{\ensuremath{\chi^2_{\text{IP}}}\xspace}

\def\gsim{{~\raise.15em\hbox{$>$}\kern-.85em
          \lower.35em\hbox{$\sim$}~}\xspace}
\def\lsim{{~\raise.15em\hbox{$<$}\kern-.85em
          \lower.35em\hbox{$\sim$}~}\xspace}

\def\pt         {\ensuremath{p_{\mathrm{T}}}\xspace}

\def\ptot       {\ensuremath{p}\xspace}

\def\evtgen     {\mbox{\textsc{EvtGen}}\xspace}

\def\geant      {\mbox{\textsc{Geant4}}\xspace}

\def\photos     {\mbox{\textsc{Photos}}\xspace}

\def\pythia     {\mbox{\textsc{Pythia}}\xspace}

\def\tell1  {TELL1\xspace}
\def\ukl1   {UKL1\xspace}

\usepackage{cite} 
\usepackage{mciteplus}

\usepackage{mciteplus}
\usepackage{footmisc}
\usepackage{longtable} 
\usepackage{booktabs} 
\usepackage{rotating} 
\usepackage{lscape}
\usepackage{units}
\usepackage{overpic}
\usepackage{adjustbox}
\usepackage{subfig}
\usepackage{pbox}
\usepackage{tabularx}
\usepackage{multirow}
\usepackage{multicol}
\usepackage{listings}
\usepackage{url}

\begin{document}

\renewcommand{\thefootnote}{\fnsymbol{footnote}}
\setcounter{footnote}{1}


\begin{titlepage}
\pagenumbering{roman}

\vspace*{-1.5cm}
\centerline{\large EUROPEAN ORGANIZATION FOR NUCLEAR RESEARCH (CERN)}
\vspace*{1.5cm}
\noindent
\begin{tabular*}{\linewidth}{lc@{\extracolsep{\fill}}r@{\extracolsep{0pt}}}
\ifthenelse{\boolean{pdflatex}}
{\vspace*{-1.5cm}\mbox{\!\!\!\includegraphics[width=.14\textwidth]{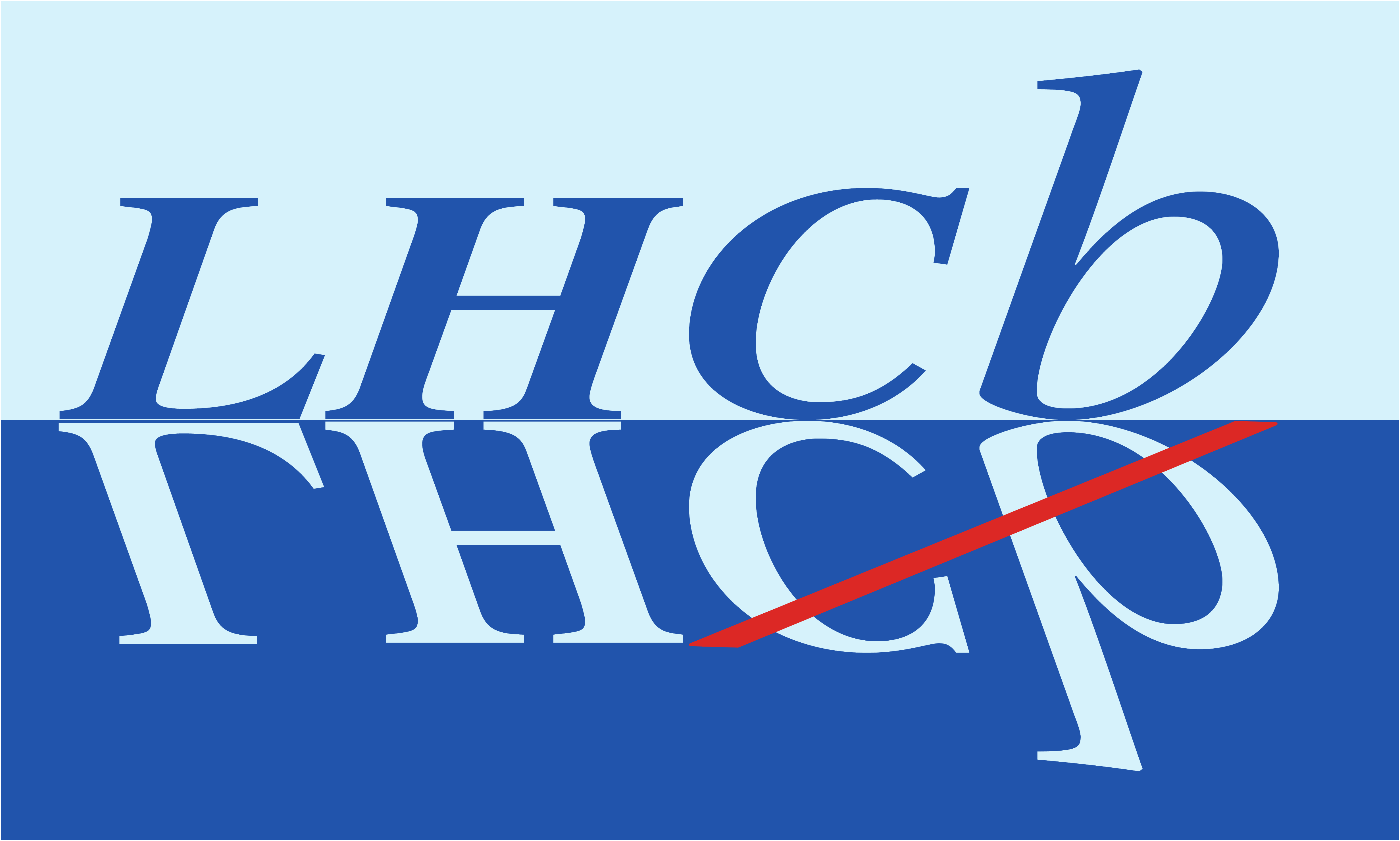}} & &}%
{\vspace*{-1.2cm}\mbox{\!\!\!\includegraphics[width=.12\textwidth]{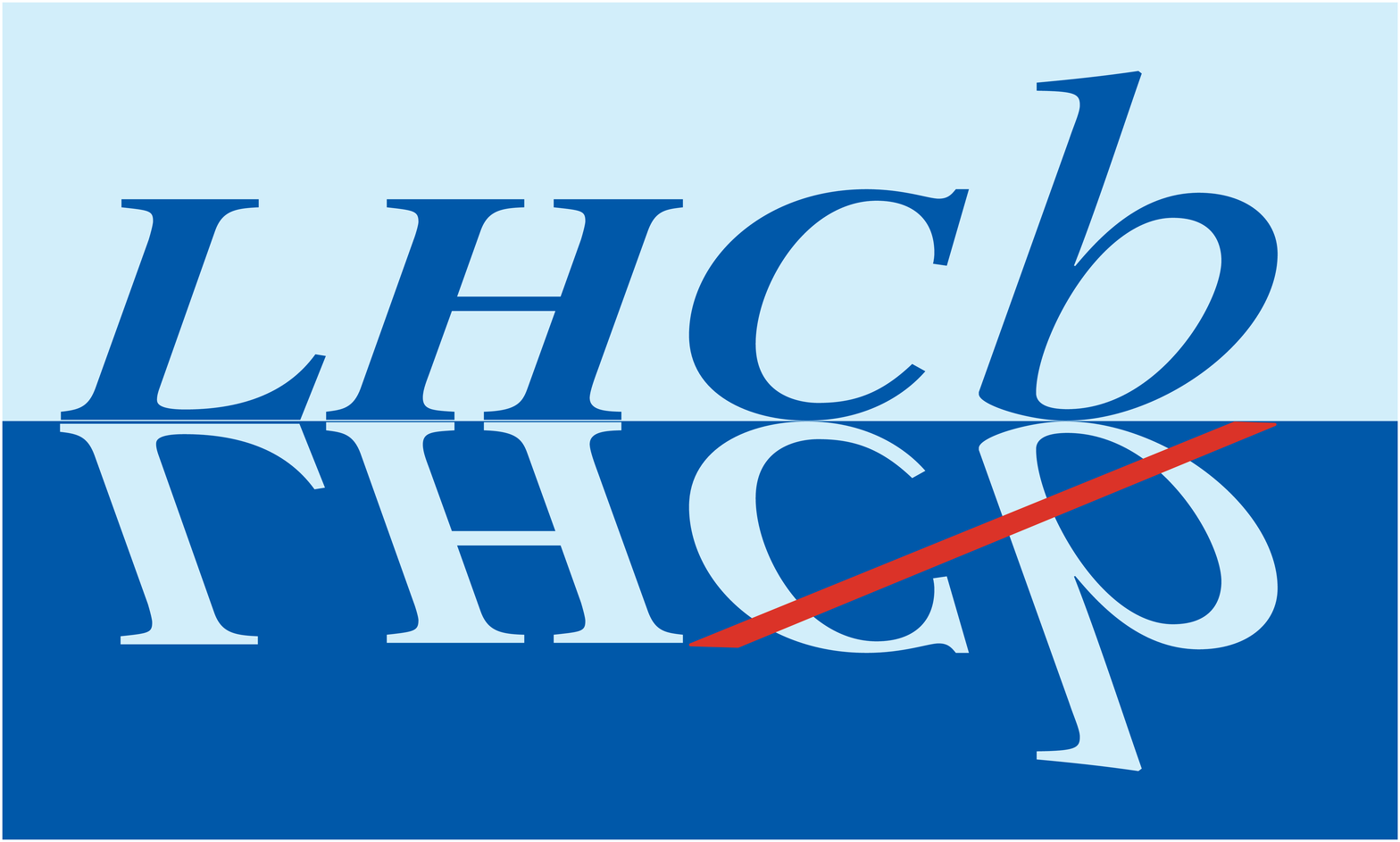}} & &}%
\\
 & & CERN-EP-2020-204 \\  
 & & LHCb-PAPER-2020-033 \\  
 & & \today \\ 
 & & \\
\end{tabular*}

\vspace*{4.0cm}

{\normalfont\bfseries\boldmath\huge
\begin{center}
  \papertitle 
\end{center}
}

\vspace*{2.0cm}

\begin{center}
\paperauthors\footnote{Authors are listed at the end of this paper.}
\end{center}

\vspace{\fill}

\begin{abstract}
  \noindent
   A search for the rare decay $\B^{0}\rightarrow\jpsi\phi$ is performed using 
   $p{\xspace}p$ collision data collected with the LHCb detector
   at centre-of-mass energies of 7, 8 and 13 \aunit{Te\kern -0.1em V}\xspace,
   corresponding to an integrated luminosity of 9 \ensuremath{\fb^{-1}}\xspace.
   No significant signal of the decay is observed and an upper limit of 
   $1.1 \times 10^{-7}$ at 90\% confidence level is set on the branching fraction.

  
\end{abstract}

\vspace*{2.0cm}

\begin{center}
  Published in Chin. Phys. C45 (2021) 043001
\end{center}

\vspace{\fill}

{\footnotesize 
\centerline{\copyright~\papercopyright. \href{\paperlicenceurl}{\paperlicence}.}}
\vspace*{2mm}

\end{titlepage}


\newpage
\setcounter{page}{2}
\mbox{~}
%
%
%
%


\renewcommand{\thefootnote}{\arabic{footnote}}
\setcounter{footnote}{0}

\cleardoublepage


\pagestyle{plain} 
\setcounter{page}{1}
\pagenumbering{arabic}


\newcommand{\les}{\ensuremath{$<$\,}\xspace}
\newcommand{\gre}{\ensuremath{$>$\,}\xspace}
\newcommand{\bel}{\ensuremath{$\in$\,}\xspace}


\newcommand{\Af}{\ensuremath{\mathcal{A}_f}\xspace}
\newcommand{\Ak}{\ensuremath{\mathcal{A}_k}\xspace}
\newcommand{\Abarf}{\ensuremath{\bar{\mathcal{A}}_f}\xspace}
\newcommand{\Abark}{\ensuremath{\bar{\mathcal{A}}_k}\xspace}
\newcommand{\Afbar}{\ensuremath{\mathcal{A}_{\bar{f}}}\xspace}
\newcommand{\Akbar}{\ensuremath{\mathcal{A}_{\bar{k}}}\xspace}
\newcommand{\Abarfbar}{\ensuremath{\bar{\mathcal{A}}_{\bar{f}}}\xspace}
\newcommand{\Abarkbar}{\ensuremath{\bar{\mathcal{A}}_{\bar{k}}}\xspace}

\def\BuToJPsiK {\decay{\Bu}{\jpsi\Kp}\xspace}

\def\BsJphi {\BsToJPsiPhi}
\def\Jmm {\jpsi(\mumu)}
\def\BsJPsimmPhi {\decay{\Bs}{\jpsi(\mumu)\phi(KK)}\xspace}
\def\BcJPsimunu {\decay{\Bc}{\Jmm \mup \Pnu}}
\newcommand{\CL}{C.L.\ }
\newcommand{\CLsb}{\ensuremath{\textrm{CL}_{\textrm{s+b}}}\xspace}
\newcommand{\CLs}{\ensuremath{\textrm{CL}_{\textrm{s}}}\xspace}
\newcommand{\CLb}{\ensuremath{\textrm{CL}_{\textrm{b}}}\xspace}
\newcommand{\MC}{Monte Carlo\xspace}

\newcommand{\bb}{\ensuremath{b\bar{b}}\xspace}
\newcommand{\mKK}{\ensuremath{m_{KK}}\xspace}
\newcommand{\mkk}{\mKK}
\newcommand{\bbdim}{\ensuremath{b\bar{b}\to \mu \mu X}\xspace}
\newcommand{\mbdim}{\ensuremath{pp\to \mu \mu X}\xspace}
\newcommand{\Bq}{\ensuremath{B^0_q}\xspace}

\newcommand{\Lambdab}{\ensuremath{\Lambda^0_b}\xspace}
\newcommand{\Bsmumu}{\ensuremath{\Bs\to\mu^+\mu^-}\xspace}
\newcommand{\Jpsimm}{\ensuremath{\Jpsi\to\mu^+\mu^-}\xspace}
\newcommand{\Bdmumu}{\ensuremath{\Bd\to\mu^+\mu^-}\xspace}
\newcommand{\KstKpi}{\ensuremath{\Kst\to K^+\pi^-}\xspace}
\newcommand{\DKpi}{\ensuremath{\D\to K^-\pi^+}\xspace}
\newcommand{\Bsmumugamma}{\ensuremath{\Bs\to\mu^+\mu^-\gamma}\xspace}
\newcommand{\Bdpipi}{\ensuremath{\Bd\to\pi^+\pi^-}\xspace}
\newcommand{\Bspipi}{\ensuremath{\Bs\to\pi^+\pi^-}\xspace}
\newcommand{\BsKK}{\ensuremath{\Bs\to K^+K^-}\xspace}
\newcommand{\BdKK}{\ensuremath{\Bd\to K^+K^-}\xspace}
\newcommand{\BdKpi}{\ensuremath{\Bd\to K^+\pi^-}\xspace}
\newcommand{\BspiK}{\ensuremath{\Bs\to\pi^+K^-}\xspace}
\newcommand{\Bhh}{\ensuremath{B^0_{(s)}\to h^+h^-}\xspace}
\newcommand{\Bmm}{\ensuremath{B^0_{(s)}\to \mu^+\mu^-}\xspace}
\newcommand{\Buhhh}{\ensuremath{B^{\pm}\to h^+h^-h^{\pm}}\xspace}
\newcommand{\BsDspi}{\ensuremath{B^0_{(s)}\to  D^-_s(K^+K^-\pi^-) \pi^+}\xspace}\newcommand{\Kst}{\ensuremath{K^{*0}}\xspace}
\newcommand{\BJpsiX}{\ensuremath{B\to J/\psi X}\xspace}
\newcommand{\BuJpsiK}{\ensuremath{B^+\to J/\psi K^+}\xspace}
\newcommand{\BuJpsimumuK}{\ensuremath{B^+\to J/\psi(\mu^+\mu^-)K^+}\xspace}
\newcommand{\BJpsimumuX}{\ensuremath{B \to J/\psi(\mu^+\mu^-)X}\xspace}
\newcommand{\BdJpsiKst}{\ensuremath{B^0_d\to J/\psi K^{*0}}\xspace}
\newcommand{\BdJpsiKpi}{\ensuremath{B^0_d\to J/\psi K \pi}\xspace}
\newcommand{\BJpsiKpi}{\ensuremath{B^0_{d(s)}\to J/\psi K \pi}\xspace}
\newcommand{\BsJpsiKpi}{\ensuremath{B^0_s\to J/\psi K \pi}\xspace}
\newcommand{\BsJpsiKst}{\ensuremath{B^0_s\to J/\psi \Kstarzb}\xspace}
\newcommand{\BJpsiKst}{\ensuremath{B^0_{d(s)}\to J/\psi K^{*0}(\Kstarzb)}\xspace}
\newcommand{\BsKstKst}{\ensuremath{\Bs\to\Kstarz\Kstarzb}\xspace}
\newcommand{\BdJpsimumuKstKpi}{\ensuremath{B^0_d\to J/\psi(\mu^+\mu^-)K^{*0}(K^+\pi^-)}\xspace}
\newcommand{\BsJpsimumuKstKpi}{\ensuremath{B^0_s\to J/\psi(\mu^+\mu^-)K^{*0}(K^+\pi^-)}\xspace}
\newcommand{\BcJpsimumumunu}{\ensuremath{B^+_c\to J/\psi(\mu^+\mu^-)\mu^+\nu_{\mu}}\xspace}
\newcommand{\Jpsimumu}{\ensuremath{J/\psi\to \mu^+\mu^-}\xspace}
\newcommand{\Jpsi}{\ensuremath{J/\psi}\xspace}

\newcommand{\Bqmumu}{\ensuremath{\ensuremath{B^0_{q}}\to\mu^+\mu^-}\xspace}
\newcommand{\BsJpsimumuPhiKK}{\ensuremath{B^0_s\to J/\psi(\mu^+\mu^-) \phi(K^+K^-)}\xspace}
\newcommand{\BdJpsimumupipi}{\ensuremath{B^0_d\to J/\psi(\mu^+\mu^-) \pi^+\pi^-}\xspace}
\newcommand{\BdJpsiRho}{\ensuremath{B^0_d\to J/\psi \rho^0}\xspace}
\newcommand{\BdJpsipipi}{\ensuremath{B^0_d\to J/\psi \pi^+\pi^-}\xspace}
\newcommand{\BsJpsiKK}{\ensuremath{B^0_s\to J/\psi K^+K^-}\xspace}
\newcommand{\ropipi}{\ensuremath{\rho^0\to \pi^+\pi^-}\xspace}

\newcommand{\Lambdappi}{\ensuremath{\Lambda\to p\pi^-}\xspace}
\newcommand{\Lambdabppi}{\ensuremath{\Lambdab\to p\pi^-}\xspace}
\newcommand{\LambdabpK}{\ensuremath{\Lambdab\to p K^-}\xspace}
\newcommand{\LbJpsipK}{\ensuremath{\Lambdab\to\Jpsi p K^-}\xspace}

\newcommand{\BsJpsiPhi}{\ensuremath{B^0_s\to J/\psi \phi}\xspace}

\newcommand{\BRof}[1]{\ensuremath{{\cal B}(#1)}\xspace}
\newcommand{\MeVc}{\ensuremath{\,{\rm MeV}/c}\xspace}
\newcommand{\GeVc}{\ensuremath{\,{\rm GeV}/c}\xspace}
\newcommand{\MeVcc}{\ensuremath{\,{\rm MeV}/c^2}\xspace}
\newcommand{\GeVcc}{\ensuremath{\,{\rm GeV}/c^2}\xspace}
\newcommand{\microb}{\ensuremath{\,{\rm \mu b}}\xspace}

\newcommand{\Enow}{\ensuremath{\,{\sqrt{s}=7\TeV}}\xspace}
\newcommand{\Einj}{\ensuremath{\,{\sqrt{s}=900\GeV}}\xspace}
\newcommand{\Enom}{\ensuremath{\,{\sqrt{s}=14\TeV}}\xspace}
\newcommand{\Efive}{\ensuremath{\,{\sqrt{s}=10\TeV}}\xspace}
\newcommand{\pdf}{\ensuremath{p.d.f.}\xspace}
\newcommand{\pdfs}{\ensuremath{p.d.f.s}\xspace}
\newcommand{\IP}{\ensuremath{{\rm IP}}\xspace}
\newcommand{\ThetaK}{\ensuremath{{\Theta_{K^{*}}}}\xspace}
\newcommand{\fdfs}{\ensuremath{\frac{f_d}{f_s}}\xspace}

\newcommand{\DLL}{\ensuremath{{\rm \Delta LL}}\xspace}
\newcommand{\gl}{\ensuremath{{\rm GL}}\xspace}
\newcommand{\glk}{\ensuremath{{\rm GL_K}}\xspace}
\newcommand{\glksm}{\ensuremath{{\rm GL_{KS}}}\xspace}
\newcommand{\glsm}{\ensuremath{{\rm GL_{S}}}\xspace}
\newcommand{\PID}{\ensuremath{{\rm PID}}\xspace}
\newcommand{\etos}{\ensuremath{\epsilon^{TOS/SEL}}\xspace}
\newcommand{\etis}{\ensuremath{\epsilon^{TIS/SEL}}\xspace}
\newcommand{\etistos}{\ensuremath{\epsilon^{TIS\&TOS/SEL}}\xspace}

\newcommand{\etrig}{\ensuremath{\epsilon^{TRIG/SEL}}\xspace}
\newcommand{\eSelect}{\ensuremath{\epsilon^{SEL/REC}}\xspace}
\newcommand{\eReco}{\ensuremath{\epsilon^{REC}}\xspace}

\newcommand{\sumpt}{\ensuremath{{\Sigma \pt}}\xspace}
\newcommand{\maxpt}{\ensuremath{{\rm max_{\pt}}}\xspace}
\newcommand{\maxip}{\ensuremath{\rm max_{\IP}}\xspace}

\newcommand{\swave}{{\em S--wave}\xspace}
\newcommand{\pwave}{{\em P--wave}\xspace}
\newcommand{\dwave}{{\em D--wave}\xspace}
\newcommand{\hwave}{{\em H--wave}\xspace}
\newcommand{\fwave}{{\em F--wave}\xspace}


\newcommand\Tstrut{\rule{0pt}{2.6ex}}
\newcommand\TTstrut{\rule{0pt}{3.2ex}}
\newcommand\Bstrut{\rule[-1.2ex]{0pt}{0pt}}
\newcommand\BBstrut{\rule[-1.8ex]{0pt}{0pt}}
\newcommand{\rhoJ}{\ensuremath{\rho_{\Jpsi}}\xspace}
\newcommand{\figref}[1]{Fig.~\ref{#1}}
\newcommand{\tabref}[1]{Table~\ref{#1}}
\newcommand{\appref}[1]{Appendix~\ref{#1}}
\newcommand{\bibref}[1]{Ref.~\cite{#1}}
\newcommand{\secref}[1]{Sect.~\ref{#1}}

\newcommand{\tred}[1]{\textcolor{red}{#1}}
\newcommand{\phisi}[1][i]{\phi_\text{s}^{#1}}
\newcommand{\phisav}{\phisi[\text{av}]}
\newcommand{\Delphisi}[1][i]{\Delta\phis^{#1}}
\newcommand{\Delphispara}{\Delphisi[\parallel]}
\newcommand{\Delphisperp}{\Delphisi[\perp]}
\newcommand{\Delphisperpp}{\Delphisi[\perp\prime]}
\newcommand{\DelphisS}{\Delphisi[\text{S}]}
\newcommand{\lamf}[1][\text{f}]{\lambda_{#1}}
\newcommand{\lamfAbs}[1][\text{f}]{|{\lamf[#1]}|}
\newcommand{\lamfSq}[1][\text{f}]{{\lamfAbs[#1]}^2}
\newcommand{\lamfbar}[1][\text{f}]{{\kern 0.06em \overline{\kern -0.06em \lambda \kern -0.03em}\kern 0.03em}_{\overline{\text{#1}}}}
\newcommand{\lamfbarAbs}[1][\text{f}]{|{\lamfbar[#1]}|}
\newcommand{\lamfbarSq}[1][\text{f}]{{\lamfbarAbs[#1]}^2}
\newcommand{\lamsAbs}{\lamfAbs[\text{s}]}
\newcommand{\lamsi}[1][i]{{\lamf[\text{s}]^{#1}}}
\newcommand{\lamsiAbs}[1][i]{|{\lamsi[#1]}|}
\newcommand{\lamsiSq}[1][i]{{\lamsiAbs[#1]}^2}
\newcommand{\Cs}{C_\text{s}}
\newcommand{\Csi}[1][i]{C_\text{s}^{#1}}
\newcommand{\Cszero}{\Csi[\text{0}]}
\newcommand{\Cspar}{\Csi[\parallel]}
\newcommand{\Csperp}{\Csi[\perp]}
\newcommand{\CsS}{\Csi[\text{S}]}
\newcommand{\Csav}{\Cs^\text{av}}
\newcommand{\DelCspara}{\Delta\Cs^\parallel}
\newcommand{\DelCsperp}{\Delta\Cs^\perp}
\newcommand{\CsavS}{\Cs^\text{avS}}
\newcommand{\taus}{\tau_\text{s}}
\newcommand{\taud}{\tau_\text{d}}
\newcommand{\mL}{M_\text{L}}
\newcommand{\mH}{M_\text{H}}
\newcommand{\Azero}[1][]{A_{\text{0}#1}}
\newcommand{\Apar}[1][]{A_{\parallel#1}}
\newcommand{\Aperp}[1][]{A_{\perp#1}}
\newcommand{\AS}[1][]{A_\text{S}#1}
\newcommand{\AAv}[1][\Ai]{#1^{\text{CP}}}
\newcommand{\AAvConj}[1][\Ai]{{\AAv[#1]}^\ast}
\newcommand{\Dms}{\Delta m_\text{s}}
\newcommand{\eGst}{e^{-\Gs\,t}}
\newcommand{\eGstal}{e^{-\Gs\,\alert{t}}}
\newcommand{\cDGs}{\cosh\!\left(\tfrac{1}{2}\DGs\, t\right)}
\newcommand{\sDGs}{\sinh\!\left(\tfrac{1}{2}\DGs\, t\right)}
\newcommand{\cDms}{\cos\!\left(\Dms\, t\right)}
\newcommand{\sDms}{\sin\!\left(\Dms\, t\right)}
\newcommand{\scDms}{\cos}
\newcommand{\ssDms}{\sin}
\newcommand{\magzero}{|\Azero|}
\newcommand{\magzeroSq}{\magzero^2}
\newcommand{\magzeroAv}{|\AAv[\Azero]|}
\newcommand{\magzeroAvSq}{\magzeroAv^2}
\newcommand{\magpar}{|\Apar|}
\newcommand{\magparSq}{\magpar^2}
\newcommand{\magparAv}{|\AAv[\Apar]|}
\newcommand{\magparAvSq}{\magparAv^2}
\newcommand{\magperp}{|\Aperp|}
\newcommand{\magperpSq}{\magperp^2}
\newcommand{\magperpAv}{|\AAv[\Aperp]|}
\newcommand{\magperpAvSq}{\magperpAv^2}
\newcommand{\magS}[1][]{|A_{\text{S}#1}|}
\newcommand{\magSSq}[1][]{\magS[#1]^2}
\newcommand{\magSAv}[1][]{|\AAv[\AS]|}
\newcommand{\magSAvSq}[1][]{\magSAv^2}
\newcommand{\FS}[1][]{F_{\text{S}#1}}
\newcommand{\FSAv}[1][]{\FS[#1]^\text{CP}}
\newcommand{\delS}[1][]{\delta_{\text{S}#1}}
\newcommand{\delparzero}{\delpar\,\text{--}\,\delzero}
\newcommand{\delperpzero}{\delperp\,\text{--}\,\delzero}
\newcommand{\delperppar}{\delperp\,\text{--}\,\delpar}
\newcommand{\delSzero}[1][]{\delS[#1]\,\text{--}\,\delzero[#1]}
\newcommand{\delSperp}[1][]{\delS[#1]\,\text{--}\,\delperp[#1]}
\newcommand{\delSpar}[1][]{\delS[#1]\,\text{--}\,\delpar[#1]}
\newcommand*\rot{\rotatebox{270}}
\newcommand{\tabcaption}[1]{  
\vspace{-\abovecaptionskip} %
\caption[#1]{#1}
\vspace{\abovecaptionskip}
}

\clearpage
\section{Introduction}
\begin{figure}[htb]
 \begin{center}
  \includegraphics[width=0.45\linewidth]{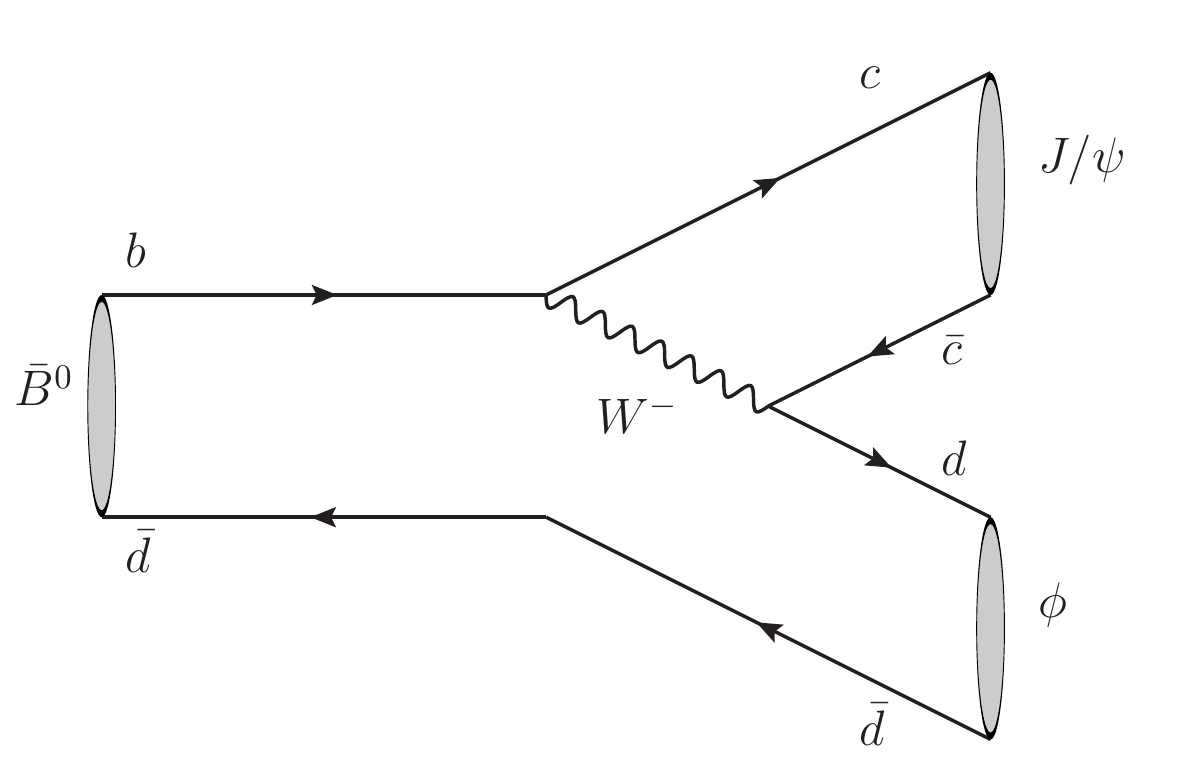}
  \put(-170,120){ (a)}
  \includegraphics[width=0.45\linewidth]{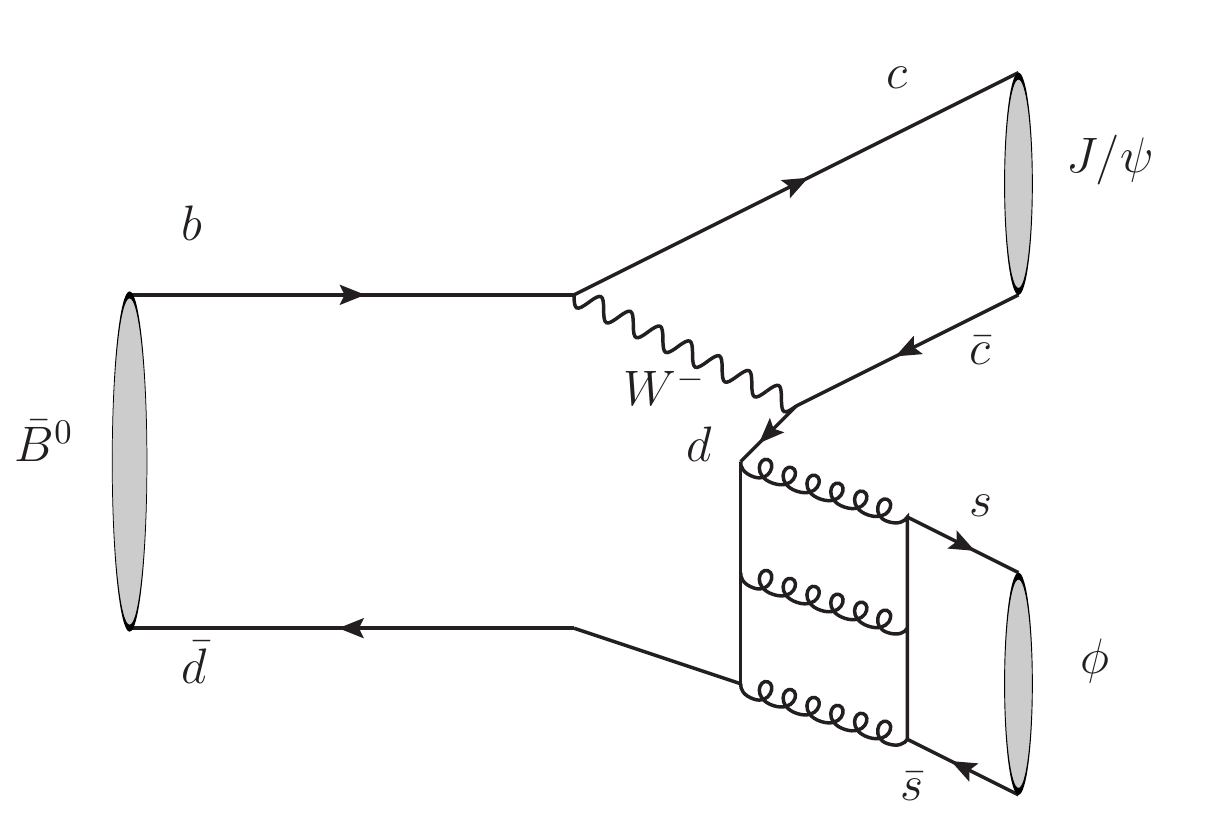}
  \put(-170,120){ (b)}
  \vspace*{-0.5cm}
 \end{center}
 \caption{ \small
  Feynman diagrams for the decay $\Bd\to\jpsi\phi$ via (a) $\omega-\phi$ mixing and (b) tri-gluon fusion.
 }
 \label{fig_feydiag}
\end{figure}

The $\Bd\to\jpsi\Kp\Km$ decay was first observed by the LHCb experiment with a branching fraction of $(2.51\pm0.35\pm0.19)\times10^{-6}$~\cite{LHCb-PAPER-2013-045}.
It proceeds primarily through the Cabibbo-suppressed $\bar{b}{\rightarrow}\bar{c}c\bar{d}$ transition. 
The $\Kp \Km$ pair can come either directly from the \Bd decay via an $s\bar{s}$ pair 
created in the vacuum, or from the decay of intermediate states that contain both $d\bar{d}$ and $s\bar{s}$
components, such as the $a_0(980)$ resonance.\footnote{The inclusion of charge-conjugate processes is implied throughout this paper.}
There is a potential contribution from the $\phi$ meson as an intermediate state. 
The decay $\Bd\to\jpsi\phi$ is suppressed by the 
Okubo-Zweig-Iizuka (OZI) rule that forbids disconnected quark diagrams~\cite{Okubo:1963fa,Zweig:570209-1,10.1143/PTPS.37.21}.
The size of this contribution and the exact mechanism to produce the $\phi$ meson in 
this process are of particular theoretical interest~\cite{Gronau:2008kk,Gronau:2008hb,Li:2009nm}. 
Under the assumption that the dominant contribution is via a small $d\bar{d}$ component in the $\phi$ wave-function, 
arising from $\omega-\phi$ mixing (Fig.~\ref{fig_feydiag}(a)), the branching fraction of the $\Bd\to\jpsi\phi$ 
decay is predicted to be of the order of $10^{-7}$~\cite{Gronau:2008kk}. 
Contributions to $\Bd\to\jpsi\phi$ decays from the OZI-suppressed tri-gluon fusion (Fig.~\ref{fig_feydiag}(b)),
photoproduction and final-state rescattering are
estimated to be at least one order of magnitude lower~\cite{Li:2009nm}.
Experimental studies of the decay $\Bd\to\jpsi\phi$ 
could provide important information about the dynamics of OZI-suppressed decays.

No significant signal of $\Bd\to\jpsi\phi$ decay has been observed in previous searches by several experiments.
Upper limits on the branching fraction of the decay have been set by 
\babar~\cite{Aubert:2003ii}, \belle~\cite{Liu:2008bta} and LHCb~\cite{LHCb-PAPER-2013-045}.
The LHCb limit was obtained using a data sample corresponding to an integrated luminosity of \mbox{1\invfb} of \proton\proton collision data, 
collected at centre-of-mass energy of 7\tev. 
This paper presents an update on the search for $\Bd\to\jpsi\phi$ decays
using a data sample corresponding to an integrated luminosity of \mbox{9\invfb},
including \mbox{3\invfb} collected at 7 and 8\tev, denoted as Run 1,
and \mbox{6\invfb} collected at 13\tev, denoted as Run 2.

The LHCb measurement in Ref.~\cite{LHCb-PAPER-2013-045} is obtained from an amplitude analysis of $\Bd\to\jpsi\Kp\Km$ decays in
a wide $m(\Kp\Km)$ range from the $\Kp\Km$ mass threshold to \mbox{2200 \mevcc}.
This paper focuses on the $\phi(1020)$ region, with the $\Kp\Km$ mass in the range \mbox{1000--1050\mevcc},
and on studies of the $\jpsi\Kp\Km$ and $\Kp\Km$ mass distributions
to distinguish the $\Bd\to\jpsi\phi$ signal
from the non-resonant decay $\Bd\to\jpsi\Kp\Km$ and background contaminations. 
The abundant decay $\Bs\to\jpsi\phi$ is used as the normalisation channel. 
The choice of mass fits over a full amplitude analysis is motivated by several considerations.
The sharp $\phi$ mass peak provides a clear signal characteristic and 
the lineshape can be very well determined using the copious $\Bs\to\jpsi\phi$ decays.
On the other hand, interference of the S-wave (either $a_0(980)/f_0(980)$ or 
non-resonant) and P-wave amplitudes vanishes in the $m(\Kp\Km)$ spectrum, 
up to negligible angular acceptance effects, 
after integrating over the angular variables.
Furthermore, significant correlations observed between $m(\jpsi\Kp\Km)$, $m(\Kp\Km)$ and angular variables
make it challenging to describe the mass-dependent angular
distributions of both  signal and  background, which are required for an amplitude analysis. 
Finally, the power of the amplitude analysis in discriminating the signal
from the non-$\phi$ contribution and background
is reduced by the large number of parameters that need to be determined in the fit.
In addition, a good understanding of the contamination from \decay{\Bs}{\jpsi\Kp\Km} decays in the $\Bd$ mass-region is essential in
the search for $\Bd\to\jpsi\phi$. 

\section{Detector and simulation }
The \lhcb detector~\cite{LHCb-DP-2008-001,LHCb-DP-2014-002} is a single-arm forward spectrometer covering 
the \mbox{pseudorapidity} range $2<\eta <5$, designed for the study of particles containing \bquark or \cquark quarks.
The detector includes a high-precision tracking system consisting of a silicon-strip vertex detector surrounding the $pp$
interaction region, a large-area silicon-strip detector located
upstream of a dipole magnet with a bending power of about
$4{\mathrm{\,Tm}}$, and three stations of silicon-strip detectors and straw
drift tubes placed downstream of the magnet.
The tracking system provides a measurement of the momentum, \ptot, of charged particles with
a relative uncertainty that varies from 0.5\% at low momentum to 1.0\% at 200\gevc.
The minimum distance of a track to a primary vertex (PV), the impact parameter (IP),
is measured with a resolution of $(15+29/\pt)\mum$,
where \pt is the component of the momentum transverse to the beam, in\,\gevc.
Different types of charged hadrons are distinguished using information
from two ring-imaging Cherenkov detectors.
Photons, electrons and hadrons are identified by a calorimeter system consisting of
scintillating-pad and preshower detectors, an electromagnetic and a hadronic calorimeter. Muons are identified by a
system composed of alternating layers of iron and multiwire proportional chambers.

Samples of simulated decays are used to optimise the signal candidate selection and derive the efficiency
of selection. In the simulation, \proton\proton collisions are generated using
\pythia~\cite{Sjostrand:2007gs,*Sjostrand:2006za} with a specific \lhcb configuration~\cite{LHCb-PROC-2010-056}.
Decays of unstable particles are described by \evtgen~\cite{Lange:2001uf}, in which final-state
radiation is generated using \photos~\cite{Golonka:2005pn}.
The interaction of the generated particles with the detector, and its response, are implemented using
the \geant toolkit~\cite{Allison:2006ve, *Agostinelli:2002hh} as described in Ref.~\cite{LHCb-PROC-2011-006}.

\section{Candidate selection}
\label{sec_evtsel}
The online event selection is performed by a trigger,
which consists of a hardware stage, based on information from the calorimeter and muon
systems, followed by a software stage, which applies a full event reconstruction.
An inclusive approach for the hardware trigger is used to maximise the available data sample, 
as described in Ref.~\cite{LHCb-PAPER-2019-013}.
Since the centre-of-mass energies and trigger thresholds are
different for the Run 1 and \mbox{Run 2} data-taking, the offline selection 
is performed separately for the two periods following the procedure described below. 
The resulting data samples for the two periods are treated separately
in the subsequent analysis procedure.

The offline selection comprises two stages. 
First, a loose selection is used to reconstruct both $\Bd\to\jpsi\phi$ and $\Bs\to\jpsi\phi$ 
candidates in the same way, given their similar kinematics.
Two oppositely charged muon candidates with $\pt > 500\mevc$ are combined to form a $\jpsi$ candidate.
The muon pair is required to have a common vertex and 
an invariant mass, $m(\mu^+\mu^-)$, in the range \mbox{3020--3170\mevcc}.
A pair of oppositely charged kaon candidates identified by the Cherenkov detectors 
is combined to form a $\phi$ candidate.
The $\Kp\Km$ pair is required to have 
an invariant mass, $m(\Kp\Km)$, in the range \mbox{1000--1050\mevcc}.
The $\jpsi$ and $\phi$ candidates are combined to form a $B^{0}_{(s)}$ candidate,
which is required to have good vertex quality and invariant mass, $m(\jpsi\Kp\Km)$,
in the range \mbox{5200--5550\mevcc}. 
The resulting $B^{0}_{(s)}$ candidate is assigned to the PV 
with which it has the smallest \chisqip, where \chisqip is defined as the difference in 
the vertex-fit \chisq of a given PV reconstructed with and without the particle being considered.
The invariant mass of the $B^{0}_{(s)}$ candidate is calculated from 
a kinematic fit for which the momentum vector of the $B^{0}_{(s)}$ candidates 
is aligned with the vector connecting the PV to the $B^{0}_{(s)}$ decay vertex and $m(\mumu)$
is constrained to the known $\jpsi$ meson mass~\cite{PDG2020}.
In order to suppress the background due to the random combination of a prompt $\jpsi$ meson and a pair of charged kaons, 
the decay time of the $B^{0}_{(s)}$ candidate is required to be greater than 0.3\ps. 

In a second selection stage, a boosted decision tree (BDT) classifier~\cite{Breiman,*AdaBoost} 
is used to further suppress combinatorial background. 
The BDT classifier is trained using simulated $\Bs\to\jpsi\phi$ decays 
representing the signal, and candidates with $m(\jpsi\Kp\Km)$ 
in the range \mbox{5480--5550\mevcc} as background.
Candidates in both samples are required to have passed the trigger and the loose selection described above.
Using a multivariate technique~\cite{multireweight}, the $\Bs\to\jpsi\phi$ simulation sample is corrected to 
match the observed distributions in background-subtracted data, including that of
the \pt and pseudorapidity of the \Bs, 
the \chisqip of the \Bs decay vertex, 
the \chisq of the decay chain of the \Bs candidate ~\cite{Hulsbergen:2005pu},
the particle identification variables, the track--fit \chisq of the muon and kaon candidates
and the numbers of tracks measured simultaneously in both the vertex detector and tracking stations.

The input variables of the BDT classifier are the minimum track--fit \chisq of the muons and the kaons,
the $\pt$ of the $B^{0}_{(s)}$ candidate and the $\Kp\Km$ combination, 
the \chisq of the $B^{0}_{(s)}$ decay vertex, 
particle identification probabilities for muons and kaons,
the minimum \chisqip of the muons and kaons, the \chisq of the $\jpsi$ decay vertex,
the \chisqip of the $B^{0}_{(s)}$ candidate and the \chisq of the $B^{0}_{(s)}$ decay chain fit.
The optimal requirement on the BDT response for the $B^{0}_{(s)}$ candidates 
is obtained by maximising the quantity $\varepsilon/\sqrt{N}$, 
where $\varepsilon$ is the signal efficiency determined in simulation and 
 $N$ is the number of candidates found in the \mbox{$\pm15\mevcc$} region
around the known \Bd mass~\cite{PDG2020}.

In addition to combinatorial background, the data also contain fake candidates from
 $\Lb\to\jpsi{p}\Km$ ($\Bd\to\jpsi\Kp\pim$) decays, where the proton (pion) is misidentified as a kaon.
To suppress these background sources, a $B^{0}_{(s)}$ candidate 
is rejected if its invariant mass, computed with one kaon interpreted as a proton (pion), 
lies within \mbox{$\pm 15\mevcc$} of the known $\Lb$ ($\Bd$) mass~\cite{PDG2020} and if the kaon candidate also satisfies proton (pion) identification requirements.

A previous study of $\Bs\to\Jpsi\phi$ decays found
that the yield of the background from $\Bd\to\Jpsi\Kp\pim$ decays 
is only 0.1\% of the $\Bs\to\jpsi\phi$ signal yield~\cite{LHCb-PAPER-2019-013}.
Furthermore, only 1.2\% of these decays, corresponding to about one candidate (three candidates) 
in the Run 1 (Run 2) data sample, fall in the \Bd mass region \mbox{5265--5295\mevcc}, according to simulation.
Thus this background is neglected.
The fraction of events containing more than one candidate 
 is 0.11\% in Run 1 data and 0.70\% in Run 2 data and these events are removed from the total data sample.
The acceptance, trigger, reconstruction and selection efficiencies of the signal 
and normalization channels are determined using simulation, 
which is corrected for the efficiency differences with respect to the data. 
The ratio of the total efficiencies of $\Bd\to\Jpsi\phi$ and $\Bs\to\Jpsi\phi$ is estimated to be
 $0.99\pm0.03\pm0.03$ for Run 1 and $0.99\pm0.01\pm0.02$ for Run 2, where
the first uncertainties are statistical and the second ones are associated with corrections to the simulation. 
The polarisation amplitudes are assumed to be the same in $\Bd\to\Jpsi\phi$ and $\Bs\to\Jpsi\phi$ decays. 
The systematic uncertainty associated with this assumption is found to be small and is neglected.

\newcommand{\invgev}{\ensuremath{\gev^{-1}}\xspace}
\newcommand{\gevt}{\ensuremath{\mathrm{Ge\kern -0.1em V}}\xspace}
\newcommand{\invgevc}{\ensuremath{{\aunit{(\gevt\!\!\!/c)}}^{-1}}\xspace}
\newcommand{\grad}{\ensuremath{^{\circ}}}
 
\section{Mass fits}
\label{sec_massfit_search}
\begin{figure}[htb]
   \begin{center}
	\includegraphics[width=0.5\linewidth]{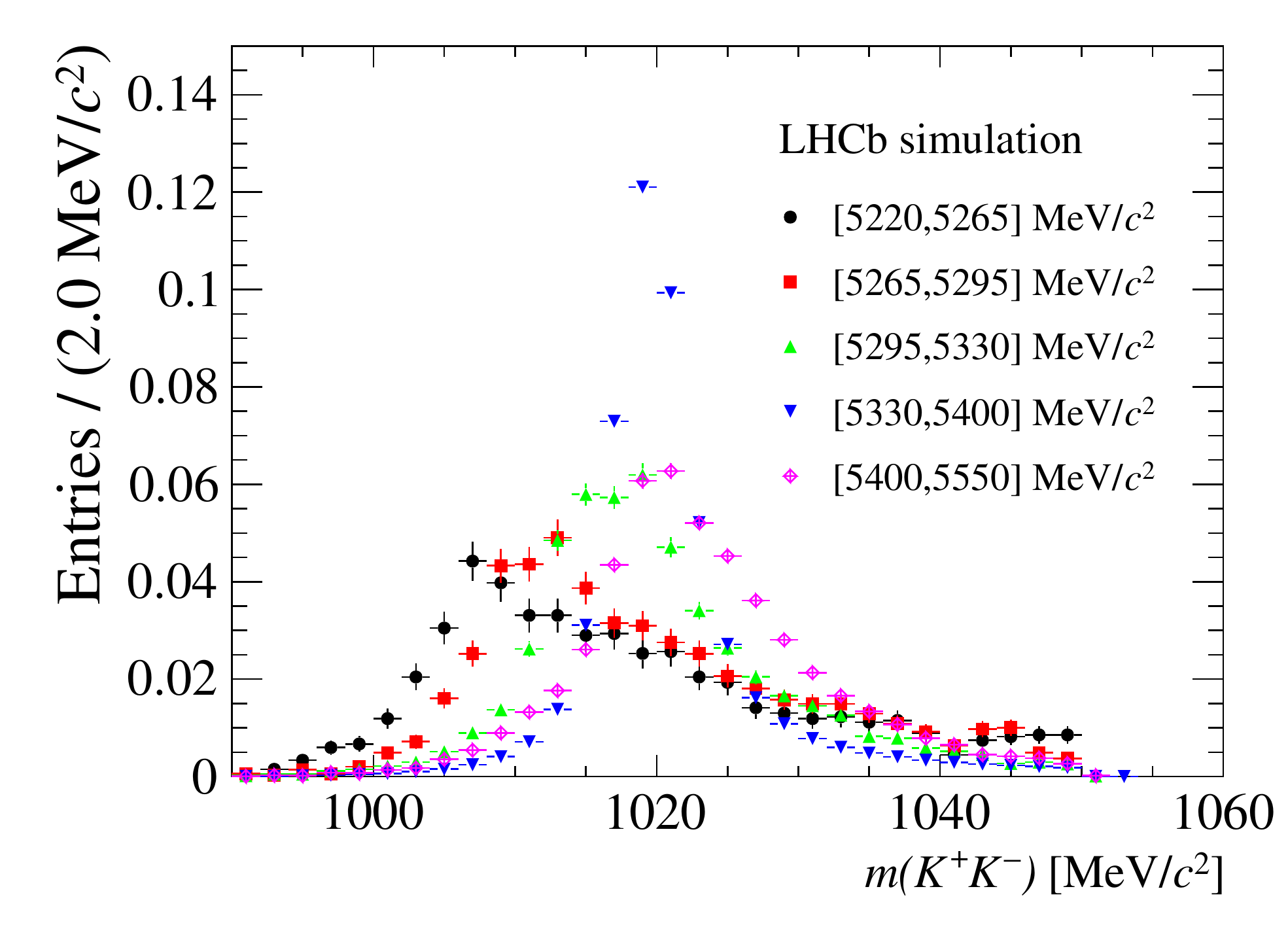}
	\vspace*{-0.5cm}
   \end{center}
   \caption{\small
	 Distributions of the invariant mass $m(\Kp\Km)$ in different $m(\jpsi\Kp\Km)$ intervals with boundaries at 5220, 5265, 5295, 5330, 5400 and 5550 \mevcc . They are obtained using simulated $\Bs\to\jpsi\phi$ decays and normalised to unity.
   }
   \label{fig_bsphi_jkk}
\end{figure}

There is a significant correlation between $m(\jpsi\Kp\Km)$ and $m(\Kp\Km)$
in $B^0_{(s)}\to\jpsi\Kp\Km$ decays, as illustrated in Fig.~\ref{fig_bsphi_jkk}, 
hence the search for $\Bd\to\jpsi\phi$ decays is carried out by performing 
sequential fits to the distributions of $m(\jpsi\Kp\Km)$ and $m(\Kp\Km)$. 
A fit to the $m(\jpsi\Kp\Km)$ distribution is used to estimate the yields of the background components
in the $\pm15\mevcc$ regions around \Bs and \Bd nominal masses.
A subsequent simultaneous fit to the $m(\Kp\Km)$ distributions of candidates 
falling in the two $\jpsi\Kp\Km$ mass windows, with the background yields 
fixed to their values from the first step,
is performed to estimate the yield of $\Bd\to\jpsi\phi$ decays.

\begin{figure}[htb]
   \begin{center}
	\includegraphics[width=0.47\linewidth]{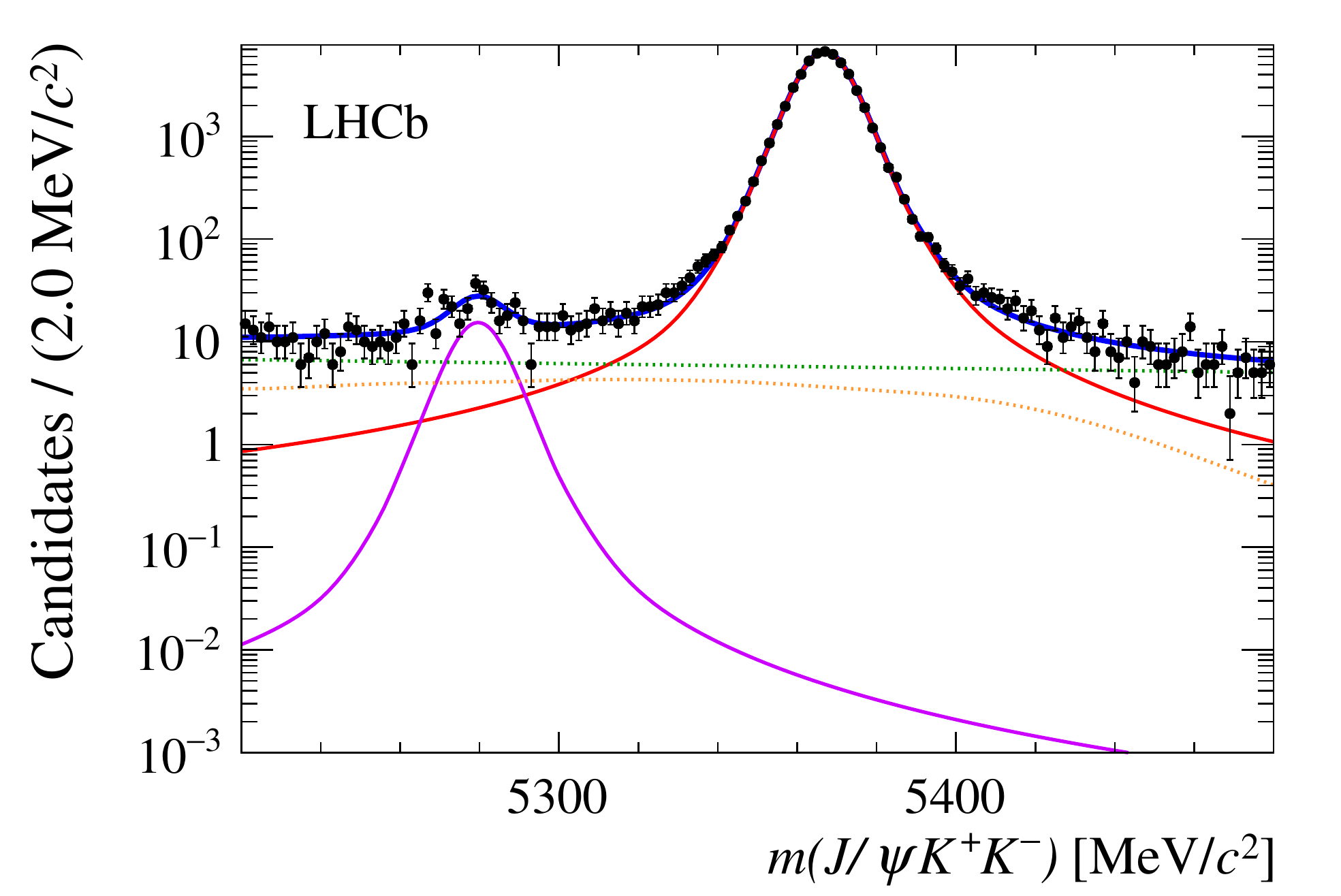}
	\put(-58,115){\small Run 1}
	\includegraphics[width=0.47\linewidth]{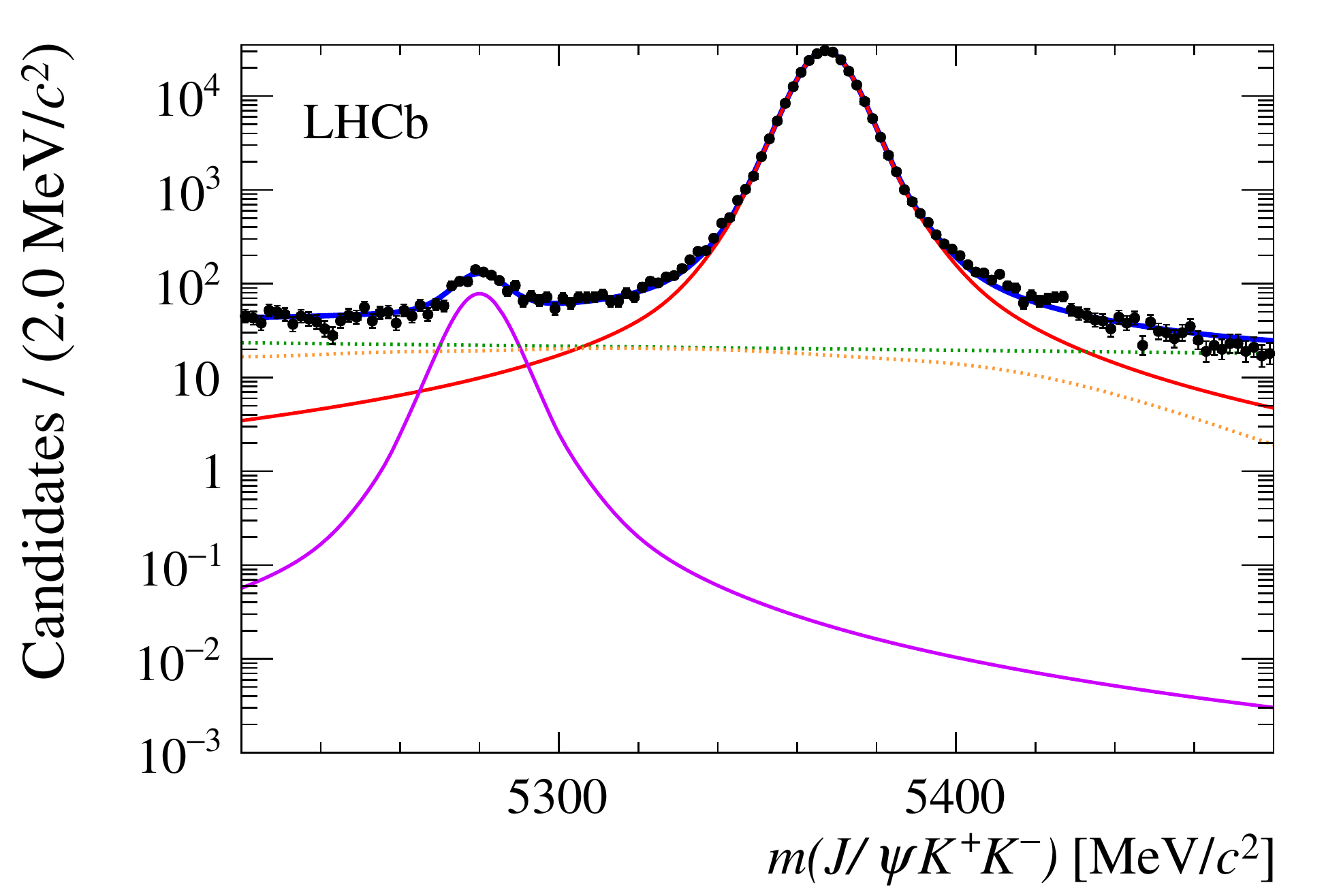}
	\put(-58,115){\small Run 2}
	
	\includegraphics[width=0.47\linewidth]{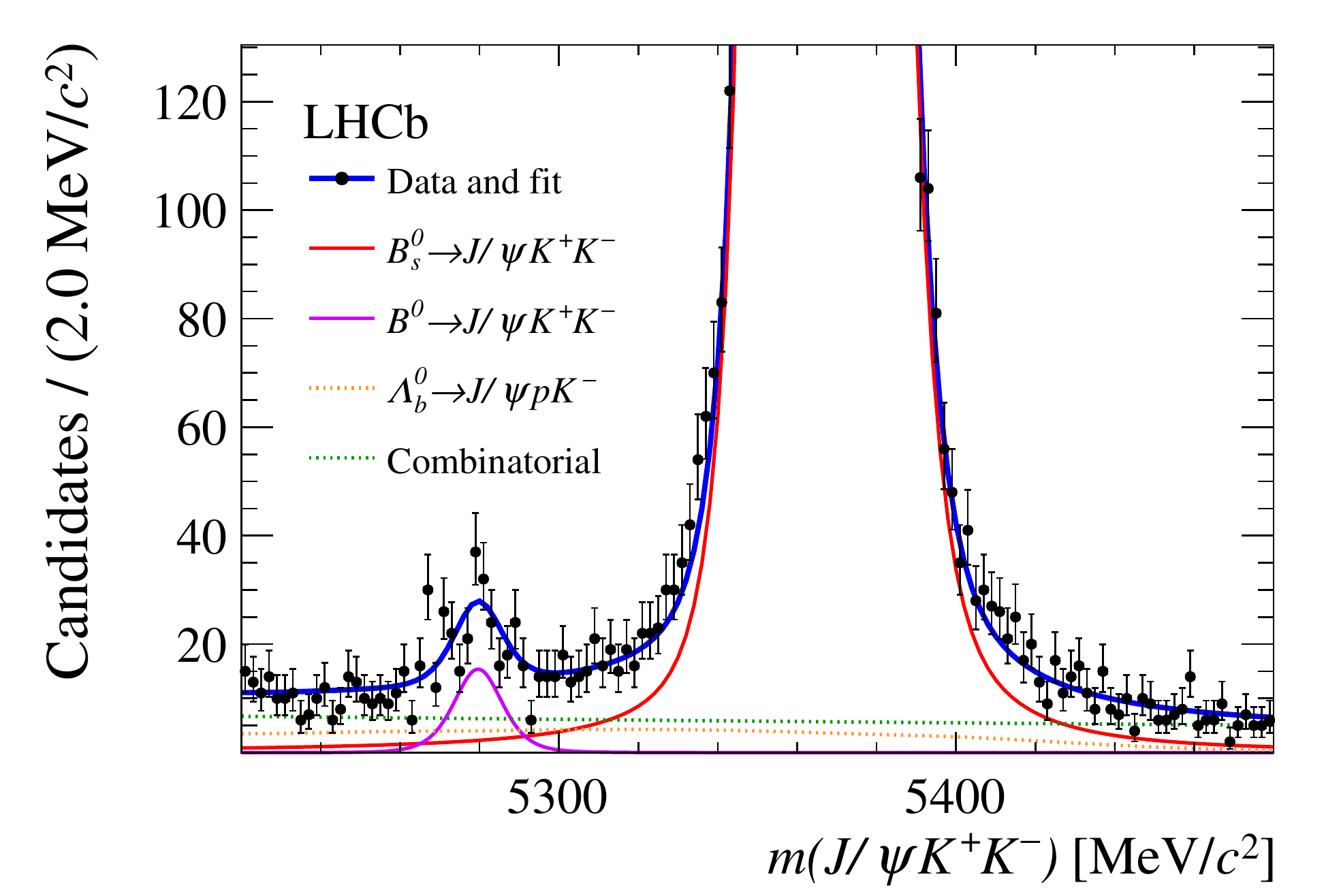}
	\put(-58,115){\small Run 1}
	\includegraphics[width=0.47\linewidth]{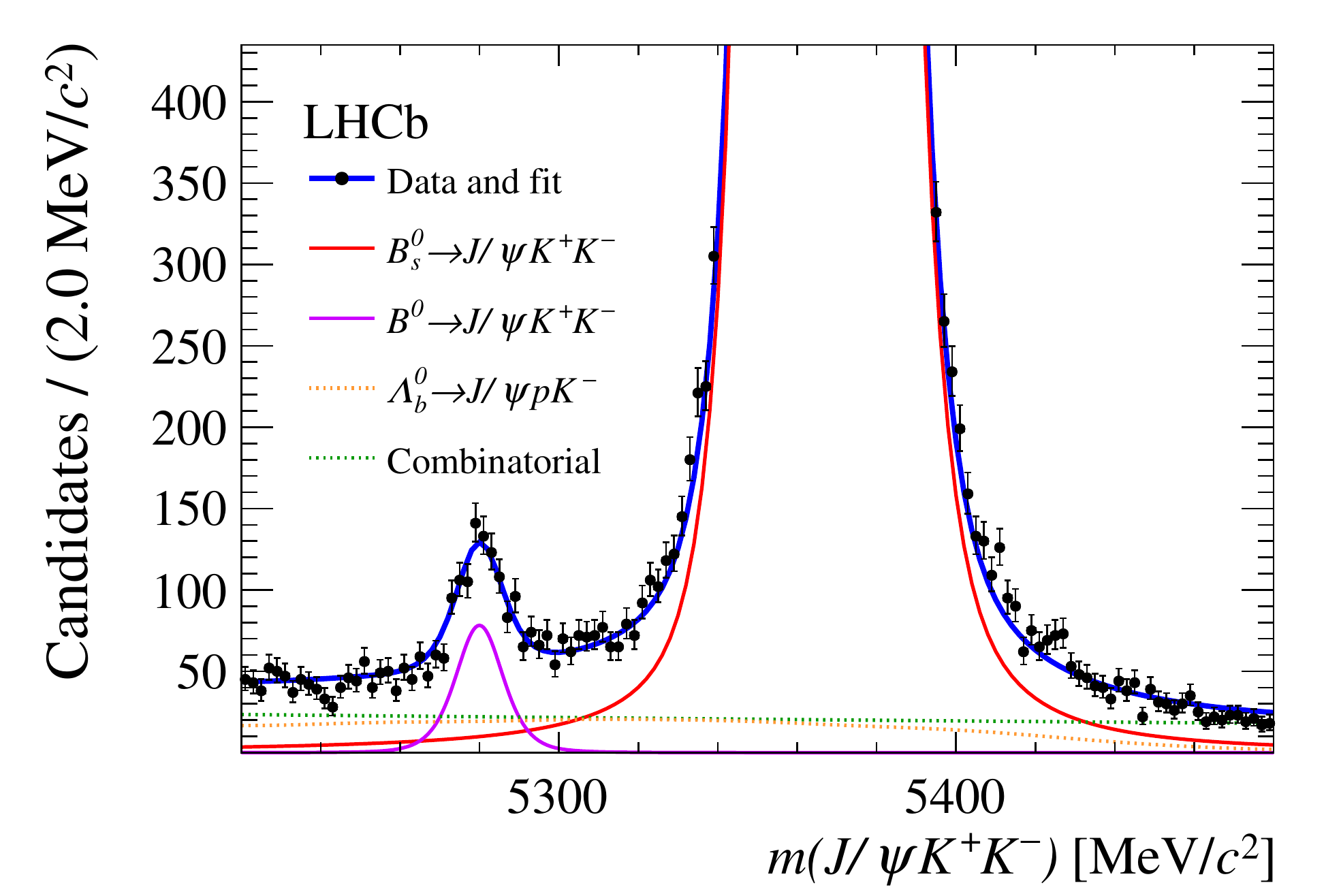}
	\put(-58,115){\small Run 2}
	
	\vspace*{-0.5cm}
   \end{center}
   \caption{\small
	The distributions of $m(\jpsi\Kp\Km)$, superimposed by the fit results,
    for (left) Run 1 and (right) Run 2 data samples, respectively.
    The top row shows the full $\Bs$ signals in logarithmic scale while
    the bottom row is presented in a reduced vertical range to make the $\Bd$ peaks visible.
	The violet (red) solid lines represent the $B^{0}_{(s)}\to\jpsi\Kp\Km$ decays,
	the orange dotted lines show the \Lb background and the green dotted lines show the combinatorial background.
   }
   \label{fig_jpsikkfit_data}
\end{figure}

The probability density function (PDF) for the $m(\jpsi\Kp\Km)$ distribution 
of both the $\Bd\to\jpsi\Kp\Km$ and $\Bs\to\jpsi\Kp\Km$ decays is modelled by the 
sum of a Hypatia ~\cite{Santos:2013gra} and a Gaussian function 
sharing the same mean. The fraction, the width ratio between the Hypatia and Gaussian functions
and the Hypatia tail parameters are determined from simulation. 
The $m(\jpsi\Kp\Km)$ shape of the $\Lb\to\jpsi{p}\Km$ background is 
described by a template obtained from simulation, while
the combinatorial background is described by an exponential function with 
the slope left to vary. 
The PDFs of $\Bd\to\jpsi\Kp\Km$ and $\Bs\to\jpsi\Kp\Km$ decays share the same shape parameters,
and the difference between the \Bs and \Bd masses is constrained to the  known mass difference of $87.23\pm0.16$ \mevcc~\cite{PDG2020}. 

An unbinned maximum-likelihood fit is performed in the $m(\jpsi\Kp\Km)$ range \mbox{5220--5480\mevcc} for Run 1 and Run 2 data samples separately. 
The yield of $\Lb\to\jpsi{p}\Km$ is estimated from a fit to the $\jpsi{p}\Km$ 
mass distribution with one kaon interpreted as a proton.
This yield is then constrained to the resulting estimate of $399\pm26$ ($1914\pm47$) 
in the $\jpsi\Kp\Km$ mass fit for the \mbox{Run 1} (\mbox{Run 2}).
The $m(\jpsi\Kp\Km)$ distributions, superimposed by the fit results, are shown in Fig.~\ref{fig_jpsikkfit_data}. 
Table~\ref{tab_sigyield} lists the obtained yields of the $\Bd\to\jpsi\Kp\Km$ and 
$\Bs\to\jpsi\Kp\Km$ decays, the \Lb background and the combinatorial background 
in the full range as well as in the $\pm15$ \mevcc regions around the known \Bs and \Bd masses.

\def\TexYa {tex/yield_tex_run1}
\def\TexYb {tex/yield_tex_run2}
\begin{table}[htb]\small
   \renewcommand\arraystretch{1.2}
   \begin{center}
      \caption{\small  Measured yields of all contributions from the fit to $\jpsi\Kp\Km$ mass distribution, showing the results for the full mass range and for the \Bs and \Bd regions. 
      }
      \label{tab_sigyield}
      \begin{tabular}{cc r@{\:$\pm$\:}l r@{\:$\pm$\:}l r@{\:$\pm$\:}l }
         \hline
         Data & Category & \multicolumn{2}{c}{Full}  & \multicolumn{2}{c}{\Bs region}  & \multicolumn{2}{c}{\Bd region} \\
         \hline
         \multirow{4}{*}{Run 1}
         & $\Bs\to\jpsi\Kp\Km$
         &55498
  &238

         &51859
 &220

         &35
 &6
\\
         
         & $\Bd\to\jpsi\Kp\Km$
         &127
  &19

         &\multicolumn{2}{c}{0
\;\;\;\;\ }
         &119
 &18
\\
         
         & $\Lb\to\jpsi{p}\Km$
         &407
     &26

         &55
 &8

         &61
 &8
\\
         
         & Combinatorial background
         &758
     &55

         &85
 &11

         &94
 &11
\\

         \hline
         \multirow{4}{*}{Run 2}
         & $\Bs\to\jpsi\Kp\Km$
         &249670
  &504

         &233663
 &472

         &153
 &12
\\

         & $\Bd\to\jpsi\Kp\Km$
         &637
  &39

         &\multicolumn{2}{c}{0
\;\;\ }
         &596
 &38
 \\

         & $\Lb\to\jpsi{p}\Km$
         &1943
     &47

         &261
 &16

         &290
 &17
\\

         & Combinatorial background
         &2677
     &109

         &303
 &20

         &331
 &21
  \\
         \hline
      \end{tabular}
   \end{center}
\end{table}

Assuming the efficiency is independent of $m(\Kp\Km)$, 
the $\phi$ meson lineshape from $\Bd\to\jpsi\phi$ ($\Bs\to\jpsi\phi$) decays 
in the \Bd (\Bs) region is given by

\begin{equation}
	{\rm S}_{\phi}(m) \equiv 
	P_{B}P_{R}F_{R}^{2}(P_{R},P_0,d)\left(\frac{P_{R}}{m^{\prime}}\right)^{2L_{R}}
	{\left|A_{\phi}(m^{\prime};m_0,\Gamma_0)\right|}^2\otimes{G(m-m^{\prime};0,\sigma)},\\
   \label{eqn_phikkpdf}
\end{equation}
where ${A}_{\phi}$ is a relativistic Breit-Wigner amplitude function~\cite{LHCb-PAPER-2015-029} defined as
\begin{equation}
	{A_{\phi}}(m;m_0,\Gamma_0) = \frac{1}{m_0^{2}-m^2-im_0\Gamma(m)}, \; 
	{\Gamma(m)} = \Gamma_0\left(\frac{P_{R}}{P_0}\right)^{2L_{R}+1}\frac{m_0}{m}F_{R}^{2}(P_{R},P_0,d) \;.
   \label{eqn_bwamp}
\end{equation}
The parameter $m$ ($m^{\prime}$) denotes the reconstructed (true) $\Kp\Km$ invariant mass,
$m_0$ and $\Gamma_0$ are the mass and decay width of the $\phi(1020)$ meson,
$P_{B}$ is the $\jpsi$ momentum in the \Bs (\Bd) rest frame,
$P_{R}$ ($P_0$) is the momentum of the kaons in the $\Kp\Km$ ($\phi(1020)$) rest frame,
$L_{R}$ is the orbital angular momentum between the $\Kp$ and $\Km$,  
$F_{R}$ is the Blatt-Weisskopf function and $d$ is the size of decaying particle, 
which is set to be \mbox{1.5$\invgevc\sim$ 0.3 fm}~\cite{VonHippel:1972fg}.
The amplitude squared is folded with a Gaussian resolution function $G$.
For $L_{R}=1$, $F_R$ has the form
\begin{equation}
   \begin{aligned}
     F_{R}(P_{R},P_0,d) = \sqrt{\frac{1+(P_{0}\,d)^2}{1+(P_{R\,}d)^2}}\;,
   \end{aligned}
\label{eqn_barfactor}
\end{equation}
and depends on the momentum of the decay products $P_{R}$~\cite{LHCb-PAPER-2015-029}.

As is shown in Fig.~\ref{fig_bsphi_jkk}, due to the correlation between the reconstructed masses of $\Kp\Km$ and $\jpsi\Kp\Km$, 
the shape of the $m(\Kp\Km)$ distribution strongly depends on the chosen $m(\jpsi\Kp\Km)$ range.
The top two plots in Fig.~\ref{fig_jpsikkfit_data} show the $m(\jpsi\Kp\Km)$ distributions for Run 1 and Run 2 separately, where a small \Bd signal can be seen on the tail of a large \Bs signal. 
Therefore, it is necessary to estimate the lineshape of the $K^+K^-$ mass spectrum from $\Bs\to\jpsi\phi$ decays in the $B^0$ region.
The $m(\Kp\Km)$ distribution of the $\Bs\to\jpsi\phi$ tail leaking into 
the \Bd mass window can be effectively described by Eq.~\ref{eqn_phikkpdf} 
with modified values of $m_0$ and $\Gamma_0$, 
which are extracted from an unbinned maximum-likelihood fit to the $\Bs\to\jpsi\phi$ simulation sample. 

The non-$\phi$ $\Kp\Km$ contributions to $\Bd\to\jpsi\Kp\Km$ ($\Bs\to\jpsi\Kp\Km$) 
decays include that from $a_0$(980)~\cite{LHCb-PAPER-2013-045} ($f_0$(980)~\cite{LHCb-PAPER-2012-040}) and nonresonant $\Kp\Km$ in an S-wave configuration.
The PDF for this contribution is given by
\begin{equation}
   {\rm S}_{non}(m) \equiv
   P_{B}P_{R}{F_B}^2\left(\frac{P_B}{m_B}\right)^{2}
   {\left|A_{R}(m)\times{e^{i\delta}}+A_{NR}\right|}^2\;,
   \label{eqn_nonpdf}
\end{equation}
where $m$ is the $\Kp\Km$ invariant mass,
$m_B$ is the known $B^{0}_{(s)}$ mass~\cite{PDG2020},
$F_B$ is Blatt-Weisskopf barrier factor of the $B^{0}_{(s)}$ meson,
$A_{R}$ and $A_{NR}$ represent the resonant ($a_0$(980) or $f_0$(980)) and nonresonant amplitudes
and $\delta$ is a relative phase between them. 
The nonresonant amplitude $A_{NR}$ is modelled as a constant function.
The lineshape of the $a_0$(980) ($f_0$(980)) resonance can be
described by a ${\rm Flatt\acute{e}}$ function~\cite{Flatte:1976xv}
considering the coupled channels $\eta\piz$ ($\pi\pi$) and $\PK\PK$.
The ${\rm Flatt\acute{e}}$ functions are given by

\begin{equation}
   A_{a_0}(m) = \frac{1}{m_{R}^{2}-m^2-i(g_{\eta\pi}^2\rho_{\eta\pi}+g_{KK}^2\rho_{KK})}
\label{eqn_a980}
\end{equation}
for the $a_0$(980) resonance and 
\begin{equation}
   A_{f_0}(m) = \frac{1}{m_{R}^{2}-m^2-im_{R}(g_{\pi\pi}\rho_{\pi\pi}+g_{KK}\rho_{KK})} 
   \label{eqn_f980}
\end{equation}
for the $f_0$(980) resonance. The parameter
$m_{R}$ denotes the pole mass of the resonance for both cases.
The constants $g_{\eta\pi}$ ($g_{\pi\pi}$) and $g_{KK}$ are the coupling strengths 
of $a_0$(980) ($f_0$(980)) to the $\eta\piz$ ($\pi\pi$) and $\PK\PK$ final states, respectively. 
The $\rho$ factors are given by the Lorentz-invariant phase space

\begin{equation}
   \rho_{\pi\pi} = \frac{2}{3}\sqrt{1-\frac{4m_{\pipm}^2}{m^2}}+\frac{1}{3}\sqrt{1-\frac{4m_{\piz}^2}{m^2}}\;,
   \label{eqn_rhopipi}
\end{equation}

\begin{equation}
   \rho_{KK} = \frac{1}{2}\sqrt{1-\frac{4m_{\Kpm}^2}{m^2}}+\frac{1}{2}\sqrt{1-\frac{4m_{\Kz}^2}{m^2}}\;,
   \label{eqn_rhokk}
\end{equation}

\begin{equation}
   \rho_{\eta\pi} = 
   \sqrt{\left(1-\frac{(m_{\eta}-m_{\piz})^2}{m^2}\right)\left(1-\frac{(m_{\eta}+m_{\piz})^2}{m^2}\right)}\;.
   \label{eqn_rhoepi}
\end{equation}
The parameters for the $a_0$(980) lineshape are
$m_{R} = 0.999\pm0.002$ \GeVcc,
$g_{\eta\pi} = 0.324\pm0.015$ $\GeVcc$, and
$g_{KK}^2/g_{\eta\pi}^2 = 1.03\pm0.14$,
determined by the Crystal Barrel experiment~\cite{Abele:1998qd};
the parameters for the $f_0$(980) lineshape are
$m_{R} = 0.9399\pm0.0063$\GeVcc,
\mbox{$g_{\pi\pi} = 0.199\pm0.030$\GeVcc}, and
$g_{KK}/g_{\pi\pi} = 3.0\pm0.3$,
according to the previous \mbox{analysis} of $\Bs\to\jpsi\pip\pim$ decays~\cite{LHCb-PAPER-2012-005}.

\begin{figure}[htb]
   \begin{center}
	\includegraphics[width=0.48\linewidth]{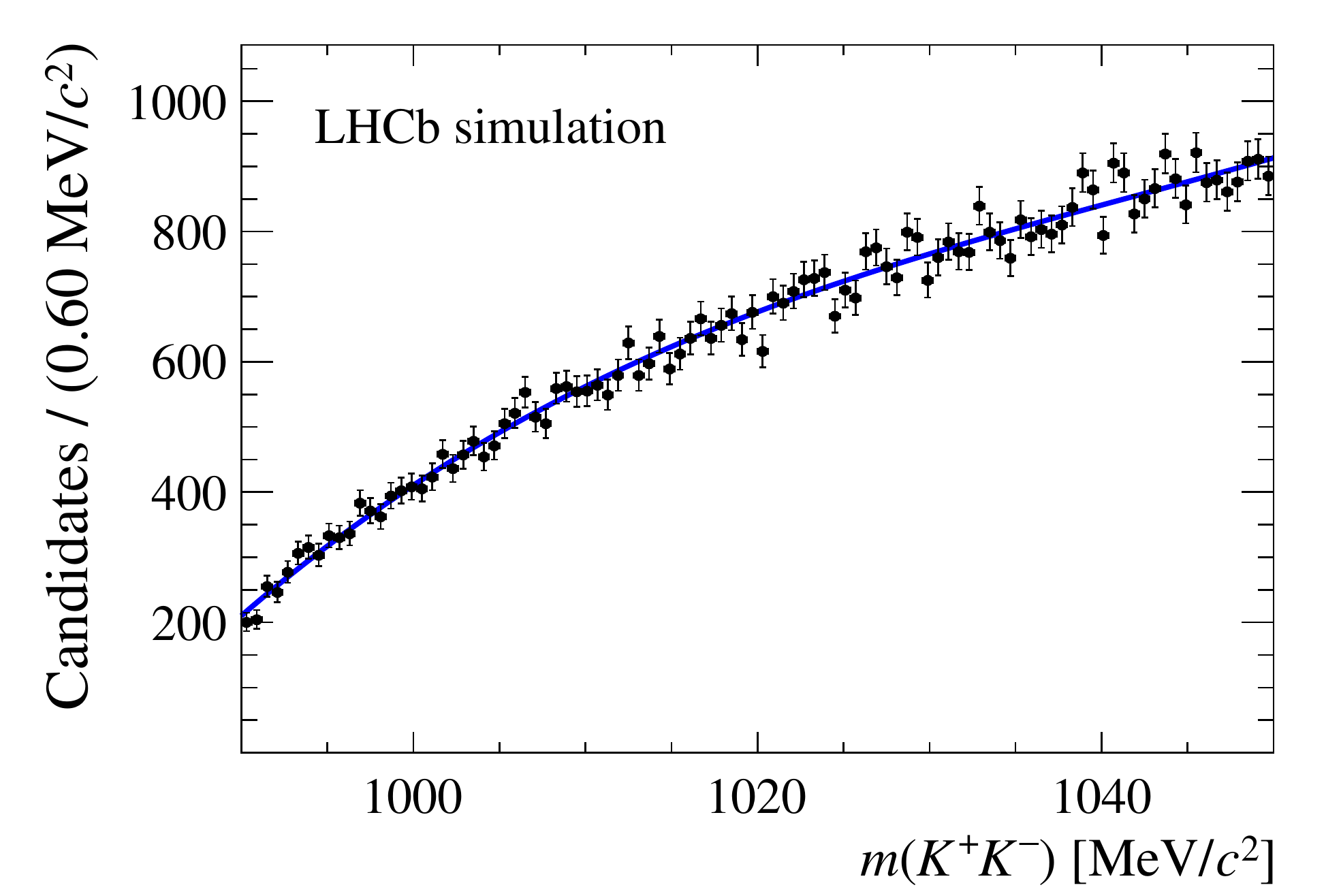}
	\vspace*{-0.5cm}
   \end{center}
   \caption{\small
	Distribution of $m(\Kp\Km)$ in a $\Lb\to\jpsi{p}\Km$ simulation sample superimposed with a fit to a polynomial function.
   }
   \label{fig_lbmkk}
\end{figure}

For the $\Lb\to\jpsi{p}\Km$ background, no dependency of the $m(\Kp\Km)$ shape on $m(\jpsi\Kp\Km)$ is observed in simulation. 
Therefore, a common PDF is used to describe the $m(\Kp\Km)$ distributions in both the \Bs and \Bd regions. 
The PDF is modelled by a third-order Chebyshev polynomial function,
obtained from the unbinned maximum-likelihood fit to the simulation shown in Fig.~\ref{fig_lbmkk}.

\begin{figure}[htb]
   \begin{center}
	\includegraphics[width=0.48\linewidth]{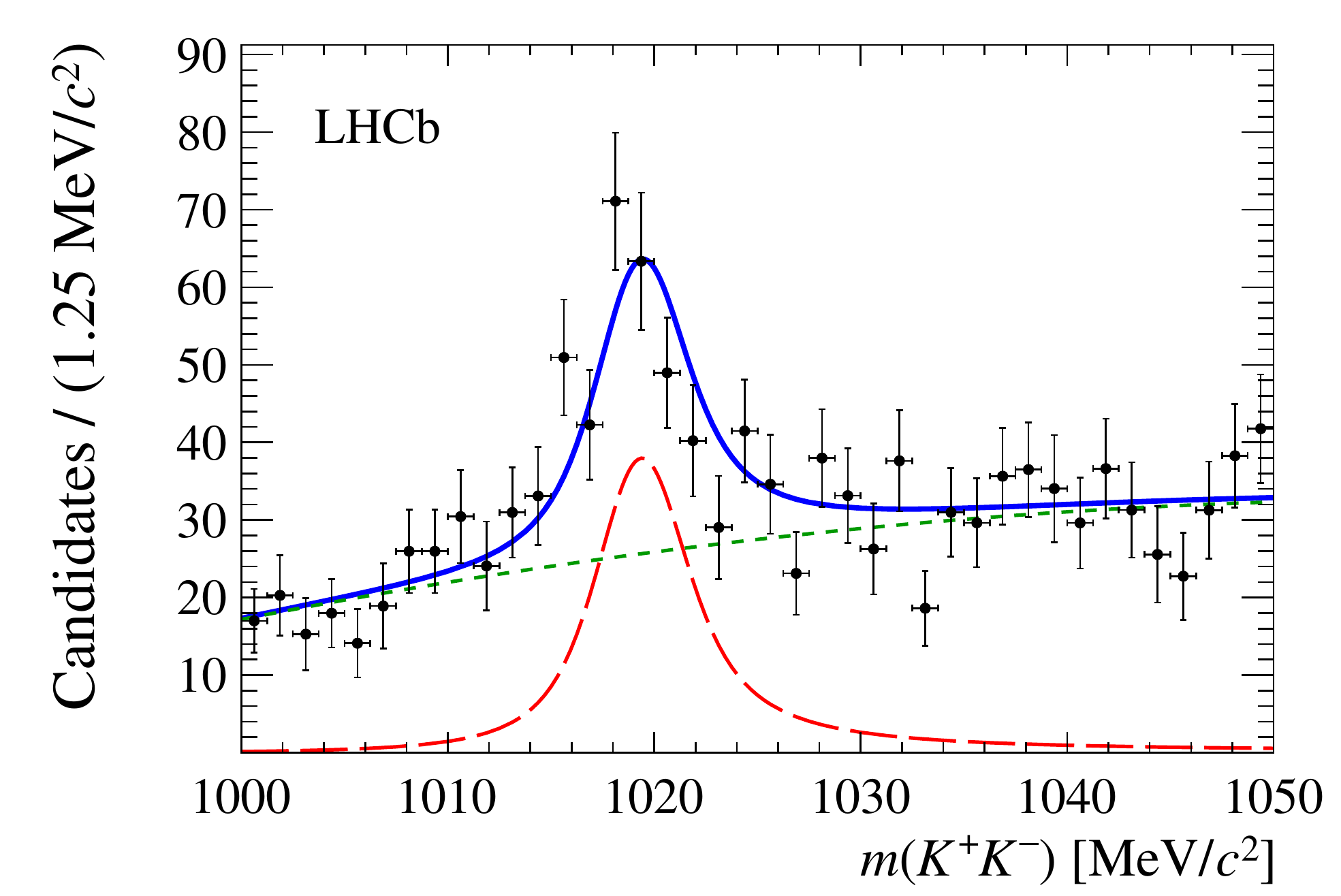}
	\put(-167,107){\small Run 1}
	\includegraphics[width=0.48\linewidth]{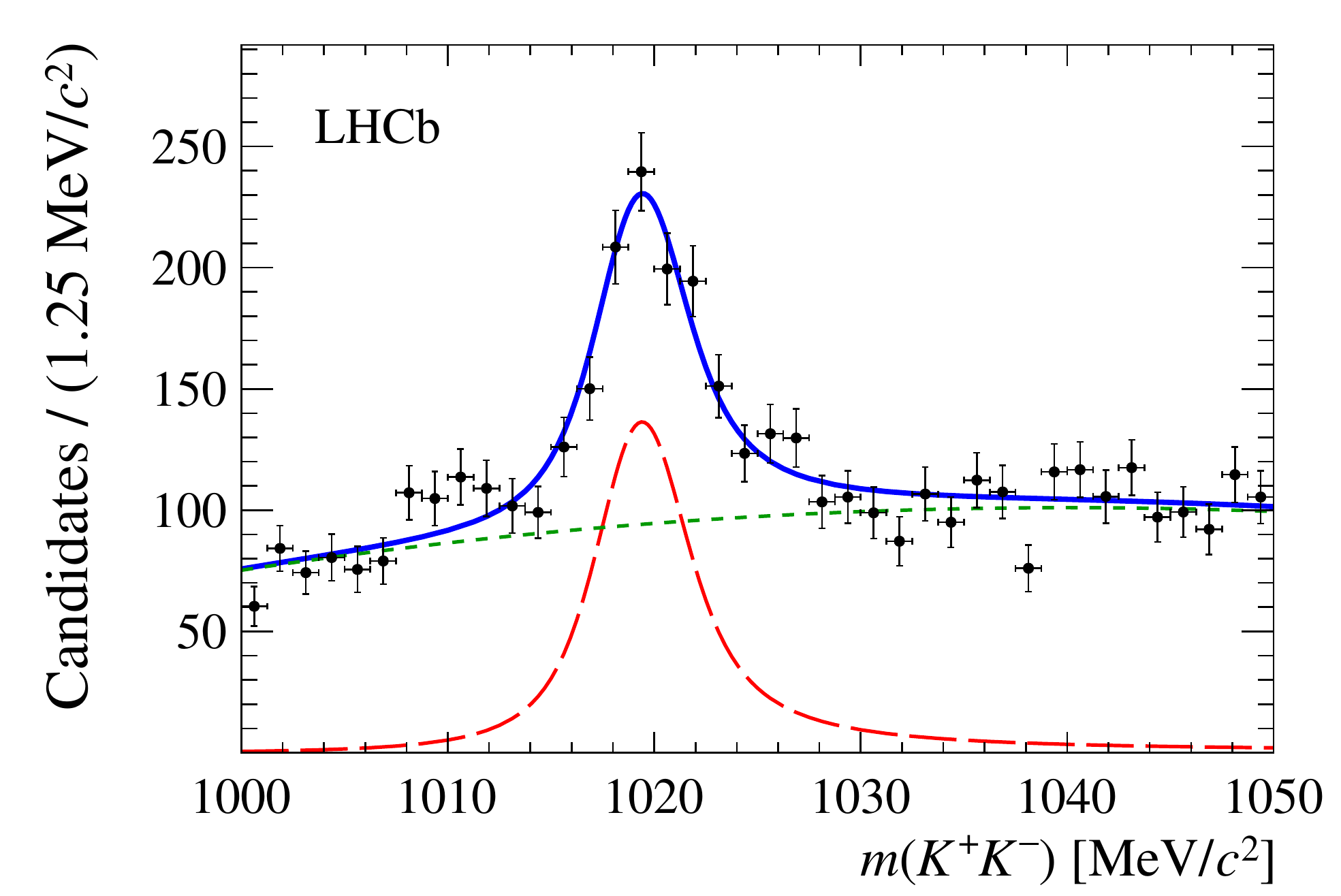}
	\put(-167,107){\small Run 2}
	\vspace*{-0.5cm}
   \end{center}
   \caption{\small
	$m(\Kp\Km)$ distributions of the enhanced combinatorial background 
    in the (left) Run 1 and (right) Run 2 data samples. 
    The $\Bs\to\jpsi\phi$ and $\Lb\to\jpsi{p}\Km$ backgrounds are subtracted by injecting simulated events with negative weights.
   }
   \label{fig_cmmkk_fit}
\end{figure}

In order to study the $m(\Kp\Km)$ shape of the combinatorial background in the \Bd region, a BDT requirement that strongly favours background is applied to form a background-dominated sample. 
Simulated $\Lb\to\jpsi{p}\Km$ and $\Bs\to\jpsi\phi$ events are then injected 
into this sample with negative weights to subtract these contributions.
The resulting $m(\Kp\Km)$ distribution is shown in Fig.~\ref{fig_cmmkk_fit}, 
which comprises a $\phi$ resonance contribution and random $\Kp\Km$ combinations, 
where the shape of the former is described by Eq.~\ref{eqn_phikkpdf}
and the latter by a second-order Chebyshev polynomial function.
To validate the underlying assumptions of this procedure, the $m(\Kp\Km)$ shape has been checked to be compatible in different $\jpsi\Kp\Km$ mass regions
and with different BDT requirements. 

A simultaneous unbinned maximum-likelihood fit to the four $m(\Kp\Km)$ distributions 
in both \Bs and \Bd regions of Run 1 and Run 2 data samples is performed. 
The $\phi$ resonance in $B^0_{(s)}\to\jpsi\phi$ decays is modelled by Eq.~\ref{eqn_phikkpdf}.
The non-$\phi$ $\Kp\Km$ contribution to $B^0_{(s)}\to\jpsi\Kp\Km $ decays is described by Eq.~\ref{eqn_nonpdf}.
The tail of $\Bs\to\jpsi\phi$ decays in the \Bd region is described by the extracted shape from simulation.
The $\Lb$ background and the combinatorial background are described by
the shapes shown in Figs.~\ref{fig_lbmkk} and~\ref{fig_cmmkk_fit}, respectively.
All $m(\Kp\Km)$ shapes are common to the \Bd and \Bs regions, 
except that of the \Bs tail, which is only needed for the \Bd region.
The mass and decay width of $\phi(1020)$ meson are constrained to their PDG values~\cite{PDG2020} while 
the width of the $m(\Kp\Km)$ resolution function is allowed to vary in the fit. 
The pole mass of $f_0$(980) ($a_0$(980)) and the coupling factors,
including $g_{\pi\pi}$, $g_{KK}/g_{\pi\pi}$, $g_{\eta\pi}^2$ and $g_{KK}^2/g_{\eta\pi}^2$,
are fixed to their central values in the reference fit.
The amplitude $A_{NR}$ is allowed to vary freely,
while the relative phase $\delta$ between the $f_0$(980) ($a_0$(980))
and nonresonance amplitudes is constrained to $-255\pm35$ ($-60\pm26$) degrees,
which was determined in the amplitude analysis of $\Bs\to\jpsi\Kp\Km$ ($\Bd\to\jpsi\Kp\Km$) decays
~\cite{LHCb-PAPER-2013-045, LHCb-PAPER-2012-040}.
The yields of the \Lb background, the $\Bs\to\jpsi\phi$ tail leaking into the \Bd region and 
combinatorial background are fixed to the corresponding values in Table~\ref{tab_sigyield},
while the yields of non-$\phi$ $\Kp\Km$ for \Bs and \Bd decays as well as 
the yield of $\Bs\to\jpsi\phi$ decays take different values 
for Run 1 and Run 2 data samples and are left to vary in the fit.

The branching fraction $\mathcal{B}(\Bd\to\jpsi\phi)$, the parameter of interest 
to be determined by the fit, is common for the Run 1 and Run 2. 
The yield of $\Bd\to\jpsi\phi$ decays is internally expressed according to 
\begin{equation}
   N_{\Bd \to \jpsi \phi} = N_{\Bs\to \jpsi \phi}\times\frac{\mathcal{B}(\Bd\to\jpsi\phi)}{\mathcal{B}(\Bs\to\jpsi\phi)}
   \times\frac{\varepsilon_{\Bd}}{\varepsilon_{\Bs}}\times\frac{1}{f_s/f_d} \;,
\label{eqn_bdsbr}
\end{equation}
where the branching fraction $\mathcal{B}(\Bs\to\jpsi\phi)$ has been measured by the LHCb collaboration~\cite{LHCb-PAPER-2012-040},
${\varepsilon_{\Bd}}/{\varepsilon_{\Bs}}$ is the efficiency ratio given in Sec.~\ref{sec_evtsel},
$f_s/f_d$ is the ratio of the production fractions of $B_s^0$ and $B^0$ mesons in \proton\proton collisions,
which has been measured at 7\tev to be $0.256\pm0.020$ in the LHCb detector acceptance~\cite{LHCb-PAPER-2012-037}. 
The effect of increasing collision energy on $f_s/f_d$ is found to be negligible for 8\tev and 
a scaling factor of $1.068\pm0.046$ is needed for 13\tev~\cite{LHCb-PAPER-2017-001}. 
The parameters $\mathcal{B}(\Bs\to\jpsi\phi)$, ${\varepsilon_{\Bd}}/{\varepsilon_{\Bs}}$ 
and $f_s/f_d$ are fixed to their central values in the baseline fit 
and their uncertainties are propagated to $\mathcal{B}(\Bd\to\jpsi\phi)$
in the evaluation of systematic uncertainties.

\begin{figure}[htb]
   \begin{center}
	\includegraphics[width=0.48\linewidth]{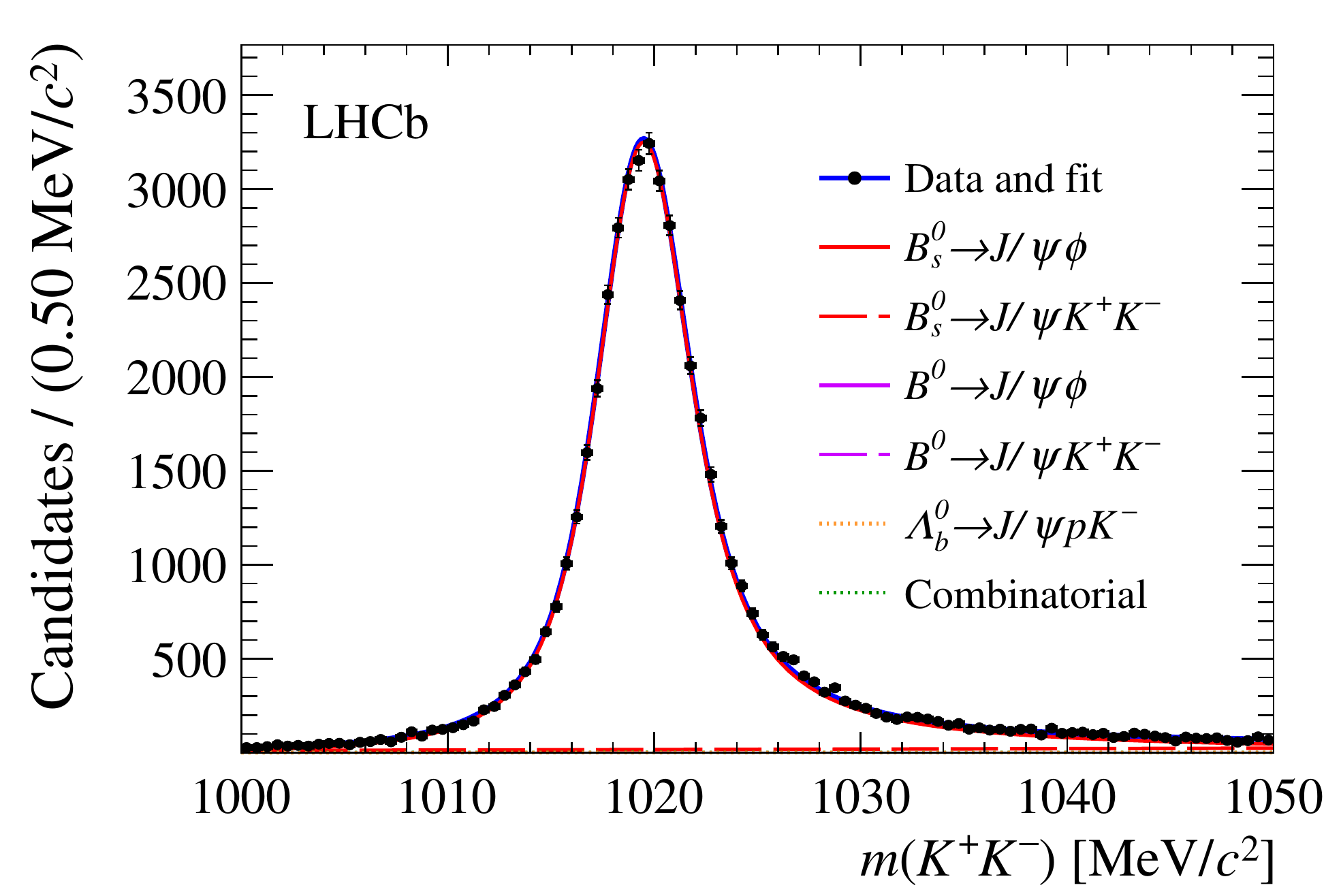}
	\put(-169,107){\small Run 1}
	\includegraphics[width=0.48\linewidth]{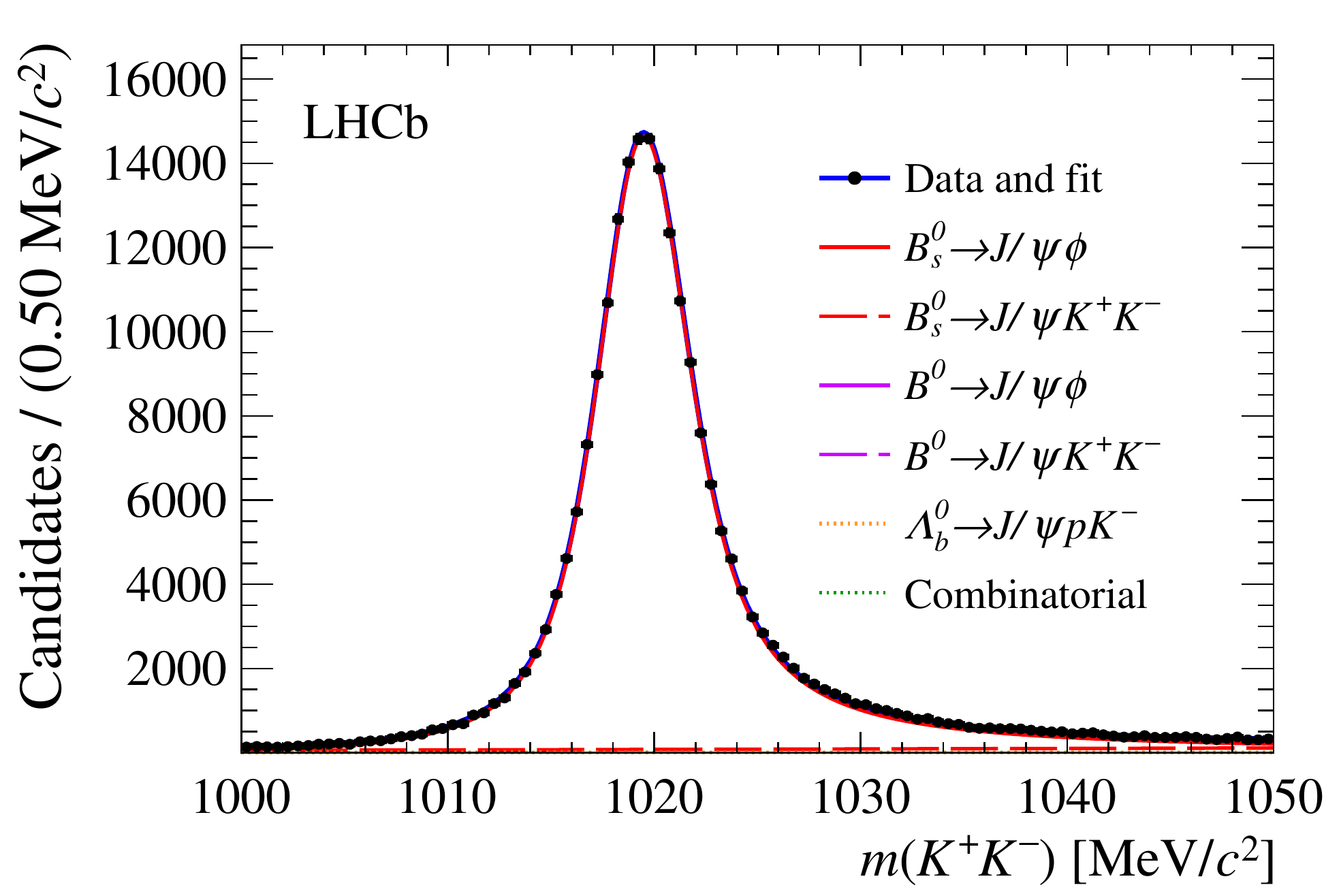}
	\put(-169,107){\small Run 2}
	
	\includegraphics[width=0.48\linewidth]{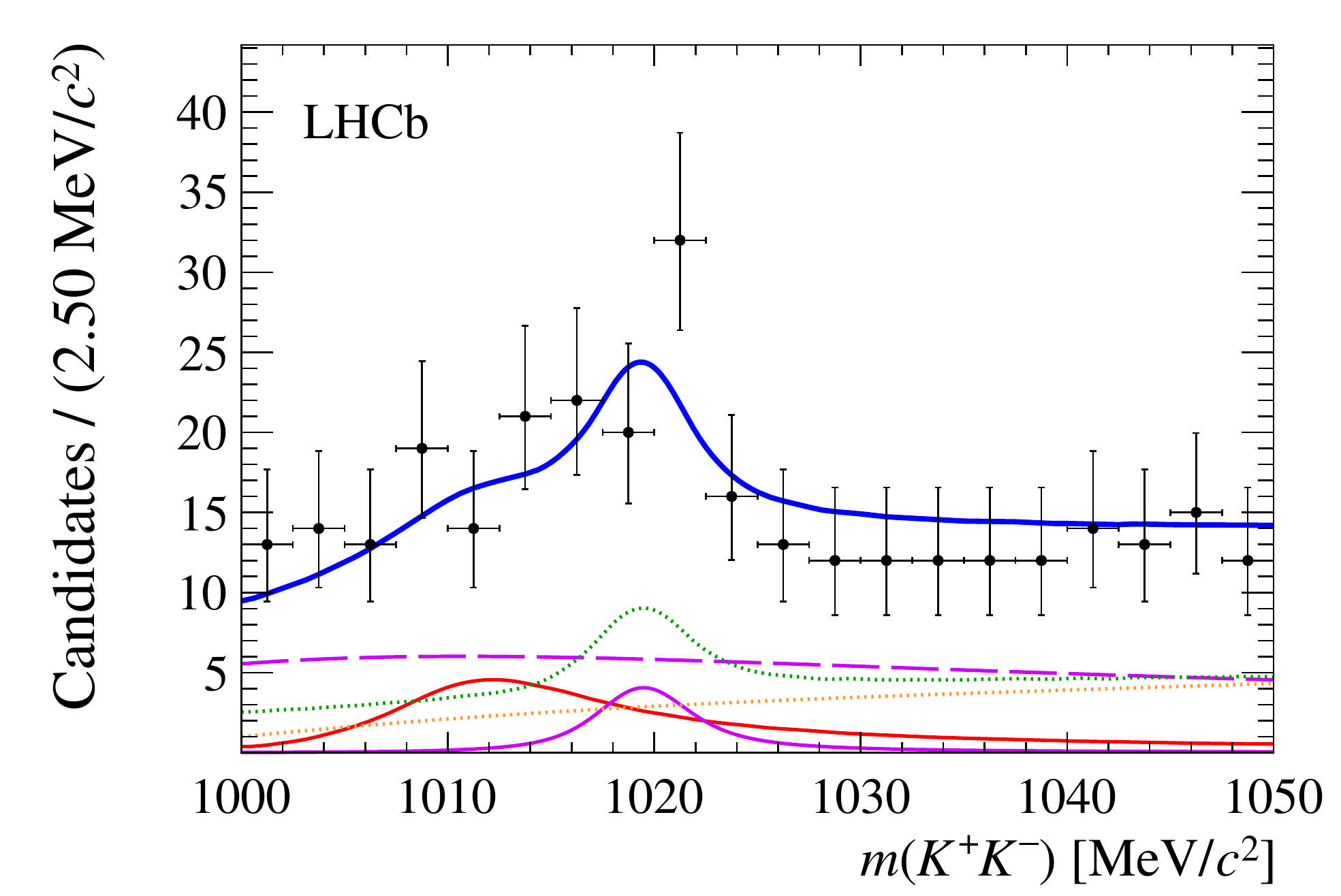}
	\put(-169,107){\small Run 1}
	\includegraphics[width=0.48\linewidth]{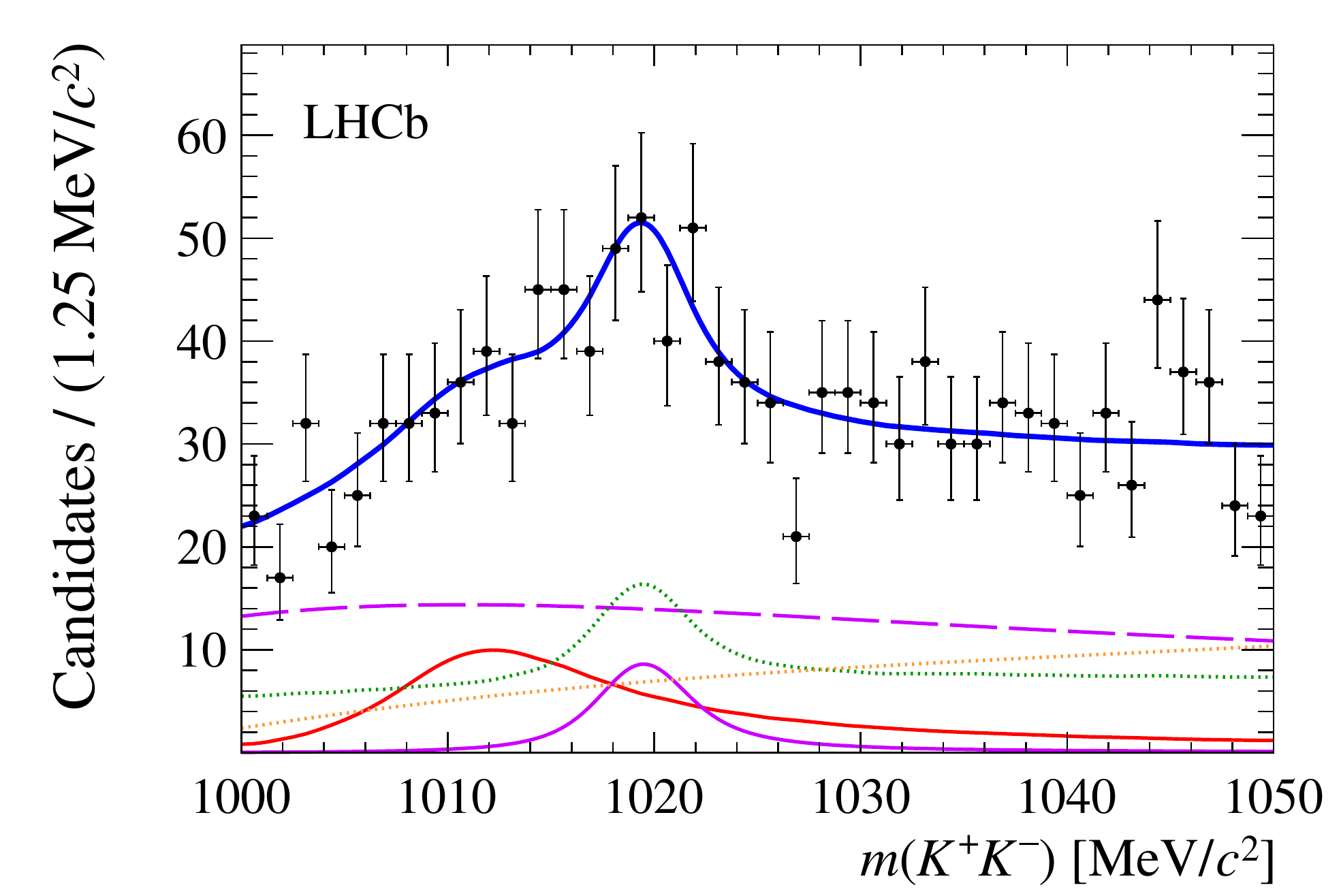}
	\put(-169,107){\small Run 2}
	\vspace*{-0.5cm}
   \end{center}
   \caption{\small
	 Distributions in the (top) \Bs and (bottom) \Bd $m(\Kp\Km)$ regions,
     superimposed by the fit results. The left and right columns
     are shown using the Run 1 and Run 2 data samples, respectively. 
     The violet (red) solid lines are $B^{0}_{(s)}\to\jpsi\phi$ decays, 
	 violet (red) dashed lines are non-$\phi$ $B_{(s)}^{0}\to\jpsi\Kp\Km$ signal,
	 green dotted lines are the combinatorial background component and 
	 the orange dotted lines are the \Lb background component.
   }
   \label{fig_totmkk_fit}
\end{figure}

The $m(\Kp\Km)$ distributions in the \Bs and \Bd regions are shown in 
Fig.~\ref{fig_totmkk_fit} for both Run 1 and Run 2 data samples. 
The branching fraction $\mathcal{B}(\Bd\to\jpsi\phi)$ 
is found to be $(6.8\pm3.0
 ({\rm stat.}))\times10^{-8}$. 
The significance of the decay $\Bd\to\jpsi\phi$, over the background-only 
hypothesis, is estimated to be 2.3
 standard deviations using Wilks' theorem~\cite{Wilks:1938dza}.

To validate the sequential fit procedure, a large number of pseudosamples are 
generated according to the fit models for the $m(J/\psi\Kp\Km)$ and $m(\Kp\Km)$ distributions.
The model parameters are taken from the result of the baseline fit to the data. 
The fit procedure described above is applied to each pseudosample. 
The distributions of the obtained estimate of $\mathcal{B}(\Bd\to\jpsi\phi)$ and 
the corresponding pulls are found to be consistent with the reference result, which indicates
that the procedure has negligible bias and its uncertainty estimate is reliable.
A similar check has been performed using pseudosamples 
generated with an alternative model for the $\Bd\to\jpsi\Kp\Km$ decays, 
which is based on the amplitude model developed for 
the $\Bs\to\jpsi\Kp\Km$ analysis~\cite{LHCb-PAPER-2019-013}
and includes contributions from P-wave $\Bd\to\jpsi\phi$ decays, 
S-wave $\Bd\to\jpsi\Kp\Km$ decays and their interference.
In this case, the robustness of the fit method has also been confirmed.

\def\TexSys {tex/systex} 
\section{Systematic uncertainties}
\label{sec_syst}

Two categories of systematic uncertainties are considered:
multiplicative uncertainties, which are associated with the normalisation factors;
and additive uncertainties, which affect the determination of the yields of the $\Bd\to\jpsi\phi$ and $\Bs\to\jpsi\phi$ modes. 

The multiplicative uncertainties include those propagated from the 
estimates of \mbox{$\mathcal{B}(\Bs\to\jpsi\phi)$}, $f_s/f_d$ and ${\varepsilon_{\Bs}}/{\varepsilon_{\Bd}}$.
Using the $f_s/f_d$ measurement at 7\tev~\cite{LHCb-PAPER-2012-040,LHCb-PAPER-2012-037},
the $\mathcal{B}(\Bs\to\jpsi\phi)$ was measured to be 
$(10.50\pm0.13\,(\rm stat.) \pm0.64\,(\rm syst.)\pm0.82\,(f_s/f_d))\times10^{-4}$. 
The third uncertainty is completely anti-correlated with the uncertainty on $f_s/f_d$, since the estimate of $\mathcal{B}(\Bs\to\jpsi\phi)$ 
is inversely proportional to the used value of $f_s/f_d$. Taking this correlation into account, yields  
$\mathcal{B}(\Bs\to\jpsi\phi)\times f_s/f_d = (2.69 \pm 0.17) \times 10^{-4}$ for 7\tev. 
The luminosity-weighted average of the scaling factor for $f_s/f_d$ for 13\tev has a relative uncertainty of 3.4\%. 
For the efficiency ratio ${\varepsilon_{\Bs}}/{\varepsilon_{\Bd}}$, 
its luminosity-weighted average has a relative uncertainty of 1.8\%.
Summing these three contributions in quadrature gives a total relative uncertainty 
of 7.3\% on $\mathcal{B}(\Bd\to\jpsi\phi)$.

The additive uncertainties are due to imperfect modeling of the
 $m(\jpsi\Kp\Km)$ and $m(\Kp\Km)$ shapes of the signal and background components.
To evaluate the systematic effect associated with the $m(\jpsi\Kp\Km)$ model 
of the combinatorial background,
the fit procedure is repeated by replacing the exponential function for the 
combinatorial background with a second-order polynomial function. 
A large number of simulated pseudosamples are generated 
according to the obtained alternative model.
Each pseudosample is fitted twice, using the baseline and 
alternative combinatorial shape, respectively. 
The average difference of $\mathcal{B}(\Bd\to\jpsi\phi)$ 
is $0.03
\times10^{-8}$, which is taken as a systematic uncertainty.

In the $m(\Kp\Km)$ fit, the yields of $\Lb\to\jpsi{p}\Km$ decay, 
combinatorial backgrounds under the \Bd and \Bs peaks 
and that of the \Bs tail leaking into the \Bd region are fixed to the values in Table~\ref{tab_sigyield}. 
Varying these yields separately leads to a change of $\mathcal{B}(\Bd\to\jpsi\phi)$
by $0.05
\times 10^{-8}$ for $\Lb\to\jpsi{p}\Km$,
$0.61
\times 10^{-8}$ for the combinatorial background and 
$0.24
\times 10^{-8}$ for the \Bs tail in the \Bd region, and these are 
assigned as systematic uncertainties on $\mathcal{B}(\Bd\to\jpsi\phi)$.

The constant $d$ in Eq.~\ref{eqn_barfactor} is varied between 1.0 and 3.0 $\invgevc$.
The maximum change of $\mathcal{B}(\Bd\to\jpsi\phi)$ is evaluated to be $0.01
\times10^{-8}$, which is taken as a systematic uncertainty.

The $m(\Kp\Km)$ shape of the \Bs tail under the \Bd peak is extracted using a $\Bs\to\jpsi\phi$ simulation sample. 
The statistical uncertainty due to the limited size of this sample
is estimated using the bootstrapping technique~\cite{efron1979}.
A large number of new data sets of the same size as the original simulation sample
are formed by randomly cloning events from the original sample, allowing one event to be cloned more than once.
The spread on the results of $\mathcal{B}(\Bd\to\jpsi\phi)$
obtained by using these pseudosamples in the analysis procedure
is then adopted as a systematic uncertainty, which is evaluated to be $0.29
\times 10^{-8}$.

In the reference model, the $m(\Kp\Km)$ shape of the $\Lb\to\jpsi{p}\Km$ background is determined from simulation,
under the assumption that this shape is insensitive to the $m(\jpsi \Kp\Km)$ region.
A sideband sample enriched with $\Lb\to\jpsi{p}\Km$ contributions
is selected by requiring one kaon to have a large probability to be a proton. 
An alternative $m(\Kp\Km)$ shape is extracted from this sample after subtracting the random combinations, 
and used in the $m(\Kp\Km)$ fit. 
The resulting change of $\mathcal{B}(\Bd\to\jpsi\phi)$ is $0.28
\times 10^{-8}$, 
which is assigned as a systematic uncertainty.

The $m(\Kp\Km)$ shape of the combinatorial background is represented by that of the $\jpsi\Kp\Km$ combinations 
with a BDT selection that strongly favours the background over the signal,
under the assumption that this shape is insensitive to the BDT requirement. 
Repeating the $m(\Kp\Km)$ fit by using the combinatorial background shape obtained with 
two non-overlapping sub-intervals of BDT response, the result of $\mathcal{B}(\Bd\to\jpsi\phi)$
is found to be stable, with a maximum variation of $0.16
\times 10^{-8}$, 
which is regarded as a systematic uncertainty.

In Equations~\ref{eqn_rhopipi}--\ref{eqn_rhoepi}, the coupling factors 
$g_{\eta\pi}$, $g_{KK}^2/g_{\eta\pi}^2$, $g_{\pi\pi}$ and $g_{KK}/g_{\pi\pi}$, 
are fixed to their mean values from Ref.~\cite{LHCb-PAPER-2012-005,Abele:1998qd}.
The fit is repeated by varying each factor by its experimental uncertainty
and the maximum variation of the branching fraction is considered for each parameter.
The sum of the variations in quadrature is $0.06
\times 10^{-8}$, 
which is assigned as a systematic uncertainty.

The systematic uncertainties are summarised in Table~\ref{tab_systs}. The total systematic uncertainty is the sum in quadrature of all these contributions. 
\begin{table}[htb]\small
   \renewcommand\arraystretch{1.2}
   \begin{center}
	\caption{\small 
	Systematic uncertainties on $\mathcal{B}(\Bd\to\jpsi\phi)$ for multiplicative and additive sources. 
	 }
	\label{tab_systs}
	\begin{tabular}{lc}
	   \hline
	  Multiplicative   uncertainties   & Value (\%) \\
	   \hline
         $\mathcal{B}(\Bs\to\jpsi\phi)$                         & $6.2$ \\
         Scaling factor for $f_{s}/f_{d}$                       & $3.4$ \\
         ${\varepsilon_{\Bd}}/{\varepsilon_{\Bs}}$              & $1.8$ \\
         \textbf{Total }                                        & $7.3$ \\
         \hline
	   Additive    uncertainties                         & Value ($10^{-8}$)\\
	   \hline
	   $m(\jpsi\Kp\Km)$ model of combinatorial background          &  \\
	   Fixed yields of $\Lb$ in $m(\Kp\Km)$ fit                    &  \\
	   Fixed yields of combinatorial background in $m(\Kp\Km)$ fit &  \\
	   Fixed yields of $\Bs$ contribution in $m(\Kp\Km)$ fit       &  \\
	   Constant $d$                                                &  \\
	   $m(\Kp\Km)$ shape of $\Bs$ contribution                     &  \\
	   $m(\Kp\Km)$ shape of $\Lb$                                  &  \\
	   $m(\Kp\Km)$ shape of combinatorial background               &  \\
	   $m(\Kp\Km)$ shape of non-$\phi$                             &  \\
	   \textbf{Total}                                              & 0.80
 \\
	   \hline
	\end{tabular}
   \end{center}
\end{table}

\begin{figure}[htb]
   \begin{center}
	\includegraphics[width=0.5\linewidth]{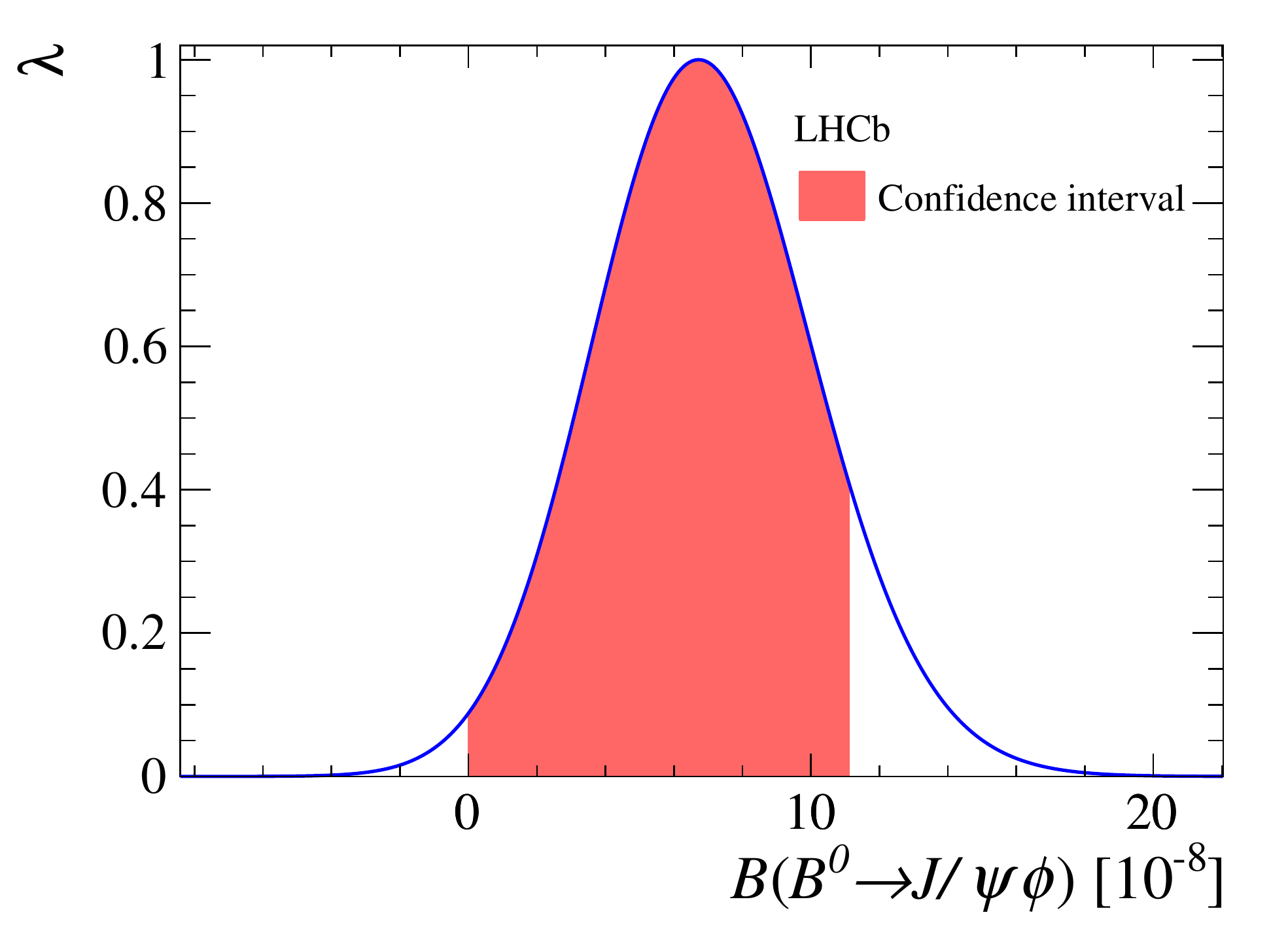}
	\vspace*{-0.5cm}
   \end{center}
   \caption{\small
	Smeared profile likelihood ratio curve shown as the blue solid line, 
    and the 90\% confidence interval indicated by the red area. 
   }
   \label{fig_smearpls}
\end{figure}

A profile likelihood method is used to compute the upper limit of 
$\mathcal{B}(\Bd\to\jpsi\phi)$~\cite{Cowan:2010js,Schott:2012zb}.
The profile likelihood ratio as a function of $ \mathcal{B} \equiv \mathcal{B}(\Bd\to\jpsi\phi)$ 
is defined as
\begin{equation}
 \lambda_0(\mathcal{B}) \equiv \frac{L(\mathcal{B},\widehat{\widehat{\nu}})}{L(\widehat{\mathcal{B}},\widehat{\nu})}\;,
\label{eqn_pls}
\end{equation}
where $\nu$ represents the set of fit parameters other than $\mathcal{B}$,
$\widehat{\mathcal{B}}$ and $\widehat{\nu}$ are the maximum likelihood estimators,
and $\widehat{\widehat{\nu}}$ is the profiled value of the parameter $\nu$ that maximises
$L$ for the specified $\mathcal{B}$.
Systematic uncertainties are incorporated by smearing the profile likelihood ratio function with 
a Gaussian function which has a zero mean and a width equal to the total systematic uncertainty:
\begin{equation}
   \lambda(\mathcal{B}) = 
   \int_{-\infty}^{+\infty}{\lambda_0}(\mathcal{B}'){\times}
   G(\mathcal{B}-\mathcal{B}',0,{\sigma_{\rm sys}}(\mathcal{B}'))d\mathcal{B}' \;.
\label{eqn_totsyst}
\end{equation}
The smeared profile likelihood ratio curve is shown in Fig.~\ref{fig_smearpls}.
The 90\% confidence interval starting at $\mathcal{B}=0$ is shown as the red area, 
which covers 90\% of the integral of the $ \lambda(\mathcal{B}) $ function in the physical region.
The obtained upper limit on $\mathcal{B}(\Bd\to\jpsi\phi)$ at 90\% CL is $1.1\times10^{-7}$.

\section{Conclusion}
\label{sec_conclusion}
A search for the rare decay $\Bd\to\jpsi\phi$ is performed using the full Run 1 and Run 2 
data samples of \proton\proton collisions collected with the LHCb experiment, 
corresponding to an integrated luminosity of 9\invfb.
A branching fraction of $\mathcal{B}(\Bd\to\jpsi\phi) = (6.8\pm3.0
\pm 0.9)\times 10^{-8}$ is measured, which indicates no statistically significant excess of the decay $\Bd\to\jpsi\phi$ above the background-only hypothesis.
The upper limit on its branching fraction at 90\% CL is determined to be $1.1\times10^{-7}$, 
which is compatible with theoretical expectations and 
improved compared with the previous limit of $1.9\times10^{-7}$ 
obtained by the LHCb experiment using Run 1 data, with a corresponding integrated luminosity of 1\invfb.

\section*{Acknowledgements}
%
%
\noindent We express our gratitude to our colleagues in the CERN
accelerator departments for the excellent performance of the LHC. We
thank the technical and administrative staff at the LHCb
institutes.
We acknowledge support from CERN and from the national agencies:
CAPES, CNPq, FAPERJ and FINEP (Brazil); 
MOST and NSFC (China); 
CNRS/IN2P3 (France); 
BMBF, DFG and MPG (Germany); 
INFN (Italy); 
NWO (Netherlands); 
MNiSW and NCN (Poland); 
MEN/IFA (Romania); 
MSHE (Russia); 
MICINN (Spain); 
SNSF and SER (Switzerland); 
NASU (Ukraine); 
STFC (United Kingdom); 
DOE NP and NSF (USA).
We acknowledge the computing resources that are provided by CERN, IN2P3
(France), KIT and DESY (Germany), INFN (Italy), SURF (Netherlands),
PIC (Spain), GridPP (United Kingdom), RRCKI and Yandex
LLC (Russia), CSCS (Switzerland), IFIN-HH (Romania), CBPF (Brazil),
PL-GRID (Poland) and OSC (USA).
We are indebted to the communities behind the multiple open-source
software packages on which we depend.
Individual groups or members have received support from
AvH Foundation (Germany);
EPLANET, Marie Sk\l{}odowska-Curie Actions and ERC (European Union);
A*MIDEX, ANR, Labex P2IO and OCEVU, and R\'{e}gion Auvergne-Rh\^{o}ne-Alpes (France);
Key Research Program of Frontier Sciences of CAS, CAS PIFI,
Thousand Talents Program, and Sci. \& Tech. Program of Guangzhou (China);
RFBR, RSF and Yandex LLC (Russia);
GVA, XuntaGal and GENCAT (Spain);
the Royal Society
and the Leverhulme Trust (United Kingdom).


\addcontentsline{toc}{section}{References}
\bibliographystyle{LHCb}
\bibliography{main,standard,LHCb-PAPER,LHCb-CONF,LHCb-DP,LHCb-TDR}

\newpage
\centerline
{\large\bf LHCb collaboration}
\begin
{flushleft}
\small
R.~Aaij$^{31}$,
C.~Abell{\'a}n~Beteta$^{49}$,
T.~Ackernley$^{59}$,
B.~Adeva$^{45}$,
M.~Adinolfi$^{53}$,
H.~Afsharnia$^{9}$,
C.A.~Aidala$^{84}$,
S.~Aiola$^{25}$,
Z.~Ajaltouni$^{9}$,
S.~Akar$^{64}$
% ...

{\footnotesize \it

% ...
$ ^{9}$Universit{\'e} Clermont Auvergne, CNRS/IN2P3, LPC, Clermont-Ferrand, France\\
% ...
$ ^{25}$INFN Sezione di Milano, Milano, Italy\\
% ...
$ ^{31}$Nikhef National Institute for Subatomic Physics, Amsterdam, Netherlands\\
% ...
$ ^{45}$Instituto Galego de F{\'\i}sica de Altas Enerx{\'\i}as (IGFAE), Universidade de Santiago de Compostela, Santiago de Compostela, Spain\\
% ...
$ ^{49}$Physik-Institut, Universit{\"a}t Z{\"u}rich, Z{\"u}rich, Switzerland\\
% ...
$ ^{53}$H.H. Wills Physics Laboratory, University of Bristol, Bristol, United Kingdom\\
% ...
$ ^{59}$Oliver Lodge Laboratory, University of Liverpool, Liverpool, United Kingdom\\
% ...
$ ^{64}$University of Cincinnati, Cincinnati, OH, United States\\
% ...
$ ^{67}$Syracuse University, Syracuse, NY, United States\\
% ...
$ ^{84}$University of Michigan, Ann Arbor, United States, associated to $^{67}$\\
\bigskip
% ...
\medskip
}
\end{flushleft}

\end{document}